\documentclass[useAMS,usenatbib]{mn2e}
\usepackage{graphicx}
\usepackage[fleqn]{amsmath}
\usepackage{amssymb}
\usepackage{multirow}

\def\ie{i.e.~}
\def\eg{e.g.~}
\def\HST{{\it{HST}}}

\def\nm{\,nm}
\def\arcsec{$^{\prime\prime}$}

\title[Transmission spectroscopy of HAT-P-32b]{The optical transmission spectrum of the hot Jupiter HAT-P-32b: clouds explain the absence of broad spectral features?}
\author[N. P. Gibson et al.]{
N. P. Gibson$^{1}$\thanks{E-mail: ngibson@eso.org},
S. Aigrain$^{2}$,
J. K. Barstow$^{2}$,
T. M. Evans$^{2}$,
L. N. Fletcher$^{3}$, and \newauthor
P. G. J. Irwin$^{3}$.
\smallskip
\\
$^{1}$European Southern Observatory, Karl-Schwarzschild-Str. 2, 85748 Garching bei M\"unchen, Germany\\
$^{2}$Department of Physics, University of Oxford, Denys Wilkinson Building, Keble Road, Oxford OX1 3RH, UK\\
$^{3}$Atmospheric, Oceanic and Planetary Physics, University of Oxford, Clarendon Laboratory, Oxford, OX1 3PU, UK\\
}

\begin{document}

\date{Accepted 2012 October 29}

\pagerange{\pageref{firstpage}--\pageref{lastpage}} \pubyear{2002}

\maketitle

\label{firstpage}

\begin{abstract}
We report Gemini-North GMOS observations of the inflated hot Jupiter HAT-P-32b during two primary transits. We simultaneously observed two comparison stars and used differential spectro-photometry to produce multi-wavelength light curves. `White' light curves and 29 `spectral' light curves were extracted for each transit and analysed to refine the system parameters and produce transmission spectra from $520-930$\,\nm\ in $\approx$14\,\nm\ bins. The light curves contain time-varying white noise as well as time-correlated noise, and we used a Gaussian process model to fit this complex noise model. Common mode corrections derived from the white light curve fits were applied to the spectral light curves which significantly improved our precision, reaching typical uncertainties in the transit depth of $\sim$$2\times10^{-4}$, corresponding to about half a pressure scale height. The low resolution transmission spectra are consistent with a featureless model, and we can confidently rule out broad features larger than about one scale height. The absence of Na/K wings or prominent TiO/VO features is most easily explained by grey absorption from clouds in the upper atmosphere, masking the spectral features. However, we cannot confidently rule out clear atmosphere models with low abundances ($\sim10^{-3}$ solar) of TiO, VO or even metal hydrides masking the Na and K wings. A smaller scale height or ionisation could also contribute to muted spectral features, but alone are unable to to account for the absence of features reported here.
\end{abstract}

\begin{keywords}
methods: data analysis, stars: individual (HAT-P-32), planetary systems, techniques: spectroscopic, techniques: Gaussian processes
\end{keywords}

\section{Introduction}

Remarkable progress has been achieved in understanding the diversity of extrasolar systems in our Galaxy, owing to the success of radial velocity and transit surveys. Transiting planets allow both their masses and radii to be measured, leading to bulk densities and a first order understanding of structure and composition. The next, natural step is to probe the atmospheres of extrasolar planets using spectroscopy, and thereby understand their chemical composition, energy budget and dominant physical processes. Transiting planets allow such measurements by analysis of temporal variations in the light received from the planet and star combined, rather than spatially resolving light from the planet.

Transmission spectroscopy is a measurement of the effective size of the planet as a function of wavelength during primary transit. The effective size of the planet is sensitive to the height at which the atmosphere becomes opaque to starlight, which in turn depends on the opacities in the atmosphere and is therefore wavelength dependent. Consequently, transmission spectroscopy enables us to probe the composition of planets' atmospheres \citep{Seager_2000,Brown_2001}.

Until quite recently, transmission spectroscopy was dominated by space-based observations \citep[e.g.][]{Charbonneau_2002,Pont_2008,Sing_2008,Sing_2011,Gibson_2012b,Huitson_2012,Berta_2012,Deming_2013}. However, nearly all attempts to produce transmission (or emission) spectra have to deal with instrumental systematics considerably larger than the atmospheric signature from the planet. We therefore rely on simple (and often arbitrary) models of the systematics to extract spectra from the time-series. Experience with the \HST/NICMOS camera shows we should be cautious when interpreting such data \citep{Gibson_2011}; indeed, repeated observations of the same target have confirmed the unreliability of such simple instrument models \citep{Crouzet_2012,Deming_2013}. Whilst other instruments have proved to be considerably more reliable \citep[\eg][]{Pont_2008,Sing_2011,Berta_2012, Deming_2013}, and better analysis techniques can remove the dependance on arbitrary systematics models \citep{Gibson_2012,Waldmann_2012}, we should remain cautious in over-interpreting marginal results from single epoch observations, as no simple systematics model can perfectly account for a complex instrument response. The best way to assess the reliability of spectra is to re-observe at the same wavelengths; unfortunately, given the expense of space based observations this is usually not feasible.

Ground-based observations of transmission spectra have been rapidly advancing in recent years with the adoption of multi-object spectrographs (MOS) to perform differential spectro-photometry \citep{Bean_2010}, and offer a powerful and complementary alternative to space based observations with some noteworthy advantages.
First, given the relative cost of observations, it is easier to obtain multiple transit observations.
Second, we can obtain continuous observations of transit light curves, unlike \HST\ which is restricted to observe in $\sim$46 minute blocks corresponding to half of its low Earth orbit.
Whilst instrumental systematics remain a problem for ground-based observations, in principle at least, this should allow us to  address the systematics more easily and therefore extract more robust  spectra. Finally, and arguably most importantly, we can access wavelength regions not obtainable using current space-based instrumentation \citep[\eg $K$-band,][]{Bean_2011}. However, ground-based differential observations are always limited by the target's nearby, bright comparison stars and therefore space based observations are likely to continue providing the strongest constraints on planetary atmospheres through observations of the brightest systems; nonetheless, ground-based observations can provide important tests on the validity of systematics models used for space-based analyses, as well as obtain reliable spectra of fainter targets.

Here we report on the use of GMOS on Gemini-North to observe the transmission spectrum of HAT-P-32b. The GMOS instruments on the Gemini telescopes have been used by \citet{Gibson_2013}, \citet{Stevenson_2013} and \citet{Crossfield_2013} to measure the transmission spectra of WASP-29b, WASP-12b and GJ 3470b, respectively, and demonstrated that a precision of $\sim1\times10^{-4}$ in transit depth is achievable. HAT-P-32b \citep{Hartman_2011} is an inflated hot-Jupiter class planet with a mass and radius of $0.860\pm0.164\,M_J$ and $1.789\pm0.025R_J$. It orbits a late-F/early-G type dwarf ($V=11.3$) with a period of 2.15\,days. Whilst our first Gemini/GMOS target, WASP-29b, has a relatively small scale height and was selected as a test case for the GMOS instrument, HAT-P-32b is larger, has a much higher equilibrium temperature ($\sim$1800\,K) and a lower surface gravity, and therefore should have a considerably larger atmospheric scale height and spectral features. Thus HAT-P-32b is the first science target in our Gemini-GMOS program.

This paper is structured as follows; in Sect.~\ref{sect:observations} we describe the observations and data reduction and in Sect.~\ref{sect:analysis} we present our light curve analysis and extraction of the transmission spectra. Finally in Sects.~\ref{sect:results} and \ref{sect:discussion} we present our results and conclusions.

\section{GMOS Observations and Data Reduction}
\label{sect:observations}

Two transits of HAT-P-32b were observed using the 8-m Gemini-North telescope with the Gemini Multi-Object Spectrograph \citep[GMOS;][]{Hook_2004} on the nights of September 5 2012 and October 18 2012. Data were taken as part of program GN-2012B-Q-16, and used a similar observing strategy to \citet{Gibson_2013}.
GMOS has an imaging field-of-view of 5.5$\times$5.5 arcmin squared, and consists of three 2048$\times$4608 pixel CCDs arranged side by side with small gaps in-between.
We observed the target ($V=11.3$, $R=11.2$) and two brighter comparison stars ($V=9.8$ and $V=11.0$) simultaneously in multi-object mode for 5.4 and 5.8 hours each night, allowing several hours either side of transit given the 3.1 hour transit duration. Conditions were not photometric for the duration of either night, and the observations were degraded due to variable cloud cover. This was considerably worse for the first transit, and we discuss the implications of this later.

\begin{figure}
\centering
\includegraphics[width=75mm]{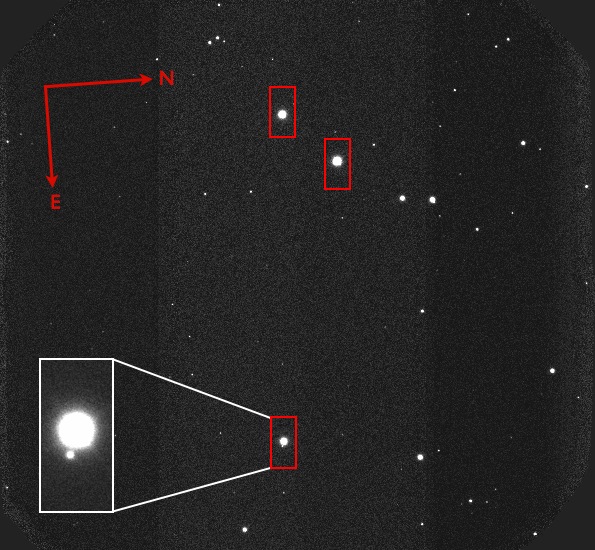}
\caption{Mosaiced pre-image of the HAT-P-32 field taken with GMOS, acquired for the mask design. The H$\alpha$ filter was used to avoid saturating the CCDs. The red boxes mark the approximate positions of the slits for the science targets, and the dispersion direction is along the x-axis. HAT-P-32 is located at the bottom of the field, and is the faintest of the three stars. The inset at the bottom left shows a zoomed in image of the target star, alongside the contaminant M-dwarf.}
\label{fig:preimage}
\end{figure}

Observations used the R400 grism + OG515 filter with a central wavelength of 725\nm\ in 2$\times$2 binning. The dispersion is 0.14\nm\ per (binned) pixel, giving wavelength coverage from about 510--930\nm. Similarly to \citet{Gibson_2013}, we read out only three regions of interest including the target and the two comparison stars to reduce the readout time to 11.5 seconds. For the first transit, exposure times started at 30 seconds and were reduced to 24 seconds towards the end of the observations to account for varying conditions, allowing for 482 exposures. For the second transit, owing to more stable conditions, the exposure times were kept at 25 seconds (except the first few exposures), resulting in 552 exposures. 
To minimise slit losses we created a mask with slits of 30\arcsec\ length and 15\arcsec\ width for the three stars designed using a pre-image taken with GMOS, giving seeing limited (therefore variable) resolution ranging from R$\approx$650--1300 at 725\nm. Fig.~\ref{fig:preimage} shows the pre-image of the field with the approximate positions of the slits marked.
Immediately before and after the observations, standard calibrations were taken consisting of flat fields and arc lamp exposures. A calibration mask was also constructed using narrower 1$^{\prime\prime}$ slits at the same positions. Arcs were taken with the calibration mask, and flat fields were obtained with both the science and calibration mask.

Data were reduced using the same procedure as \citet{Gibson_2013}, with the standard GMOS pipeline contained in the Gemini {\sc IRAF}\footnote{{\sc IRAF} is distributed by the National Optical Astronomy Observatory, which is operated by the Association of Universities for Research in Astronomy (AURA) under cooperative agreement with the National Science Foundation}/{\sc PyRAF}\footnote{{\sc PyRAF} is a product of the Space Telescope Science Institute, which is operated by AURA for NASA} package. First the ROI images were processed to be in the standard GMOS format. Basic reductions included bias subtraction and wavelength calibration. A notable difference in our GMOS-North observations compared to GMOS-South is that fringing is not present at a significant level, and consequently we were able to extract useful data over the full spectral range. We flat fielded the data using flats taken with the science mask, although we experimented with various strategies and note this had little influence on the results.

Wavelength calibrated spectra for the 3 stars were extracted by simply summing in the cross-dispersion direction after sky subtraction using the {\sc gsextract} routine, with an aperture (diameter) of 4\arcsec\ ($\approx28$ binned pixels). A few pixel columns (\ie along the spatial direction) showing significant temporal variation were masked from the extraction. The aperture size was chosen to both optimise the RMS, and to avoid a previously unknown companion star $\approx$2.8\arcsec\ ($\approx$20 binned pixels) from the target star. The location of this star is shown in Fig.~\ref{fig:preimage}. 
Whilst the contaminant spectrum is resolved, there is some overlap in the cross-dispersion PSFs. With our final extraction aperture, we estimate a few percent of the contaminant star's flux will fall within our aperture for HAT-P-32. The contaminant star is an M-dwarf, and consequently will only affect the transmission spectrum at the reddest wavelengths. At these wavelengths the detector registers about 1/60th of the peak flux from the contaminant star compared to the target. Combined with the estimated few percent of light that falls within the aperture we calculate that the contaminant could dilute the transit depth at the reddest wavelengths by less than $10^{-5}$, far below the expected precision of our observations. For this crude calculation we assumed a Gaussian PSF, and median seeing. However, even with larger contamination we expect this would not significantly dilute our light curves, but we keep this in mind during interpretation of our results, and in comparison with published parameters of the HAT-P-32 system. More detailed correction for the contaminant star is difficult, given the varying seeing over the course of the observations, although we note that our results do not change considerably when using a larger aperture.

\begin{figure}
\includegraphics[width=85mm]{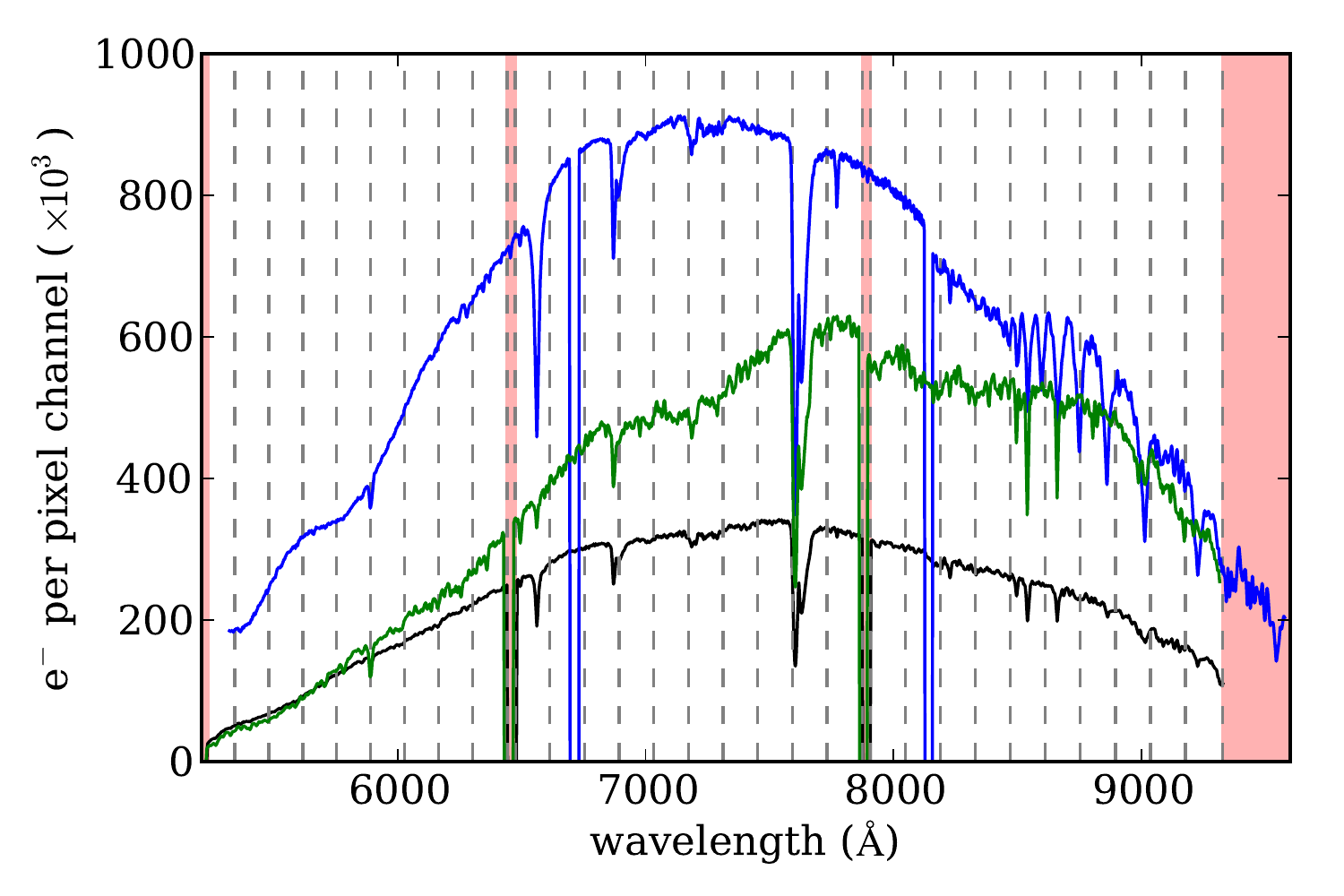}
\caption{Example spectra extracted from a single exposure. The black line is HAT-P-32, and the blue and green lines the two comparison stars. Dashed lines define the wavelength channels, and shading marks excluded regions from the analysis at the edges and in-between detectors.}
\label{fig:eg_spec}
\end{figure}

\begin{figure*}
\includegraphics[width=85mm]{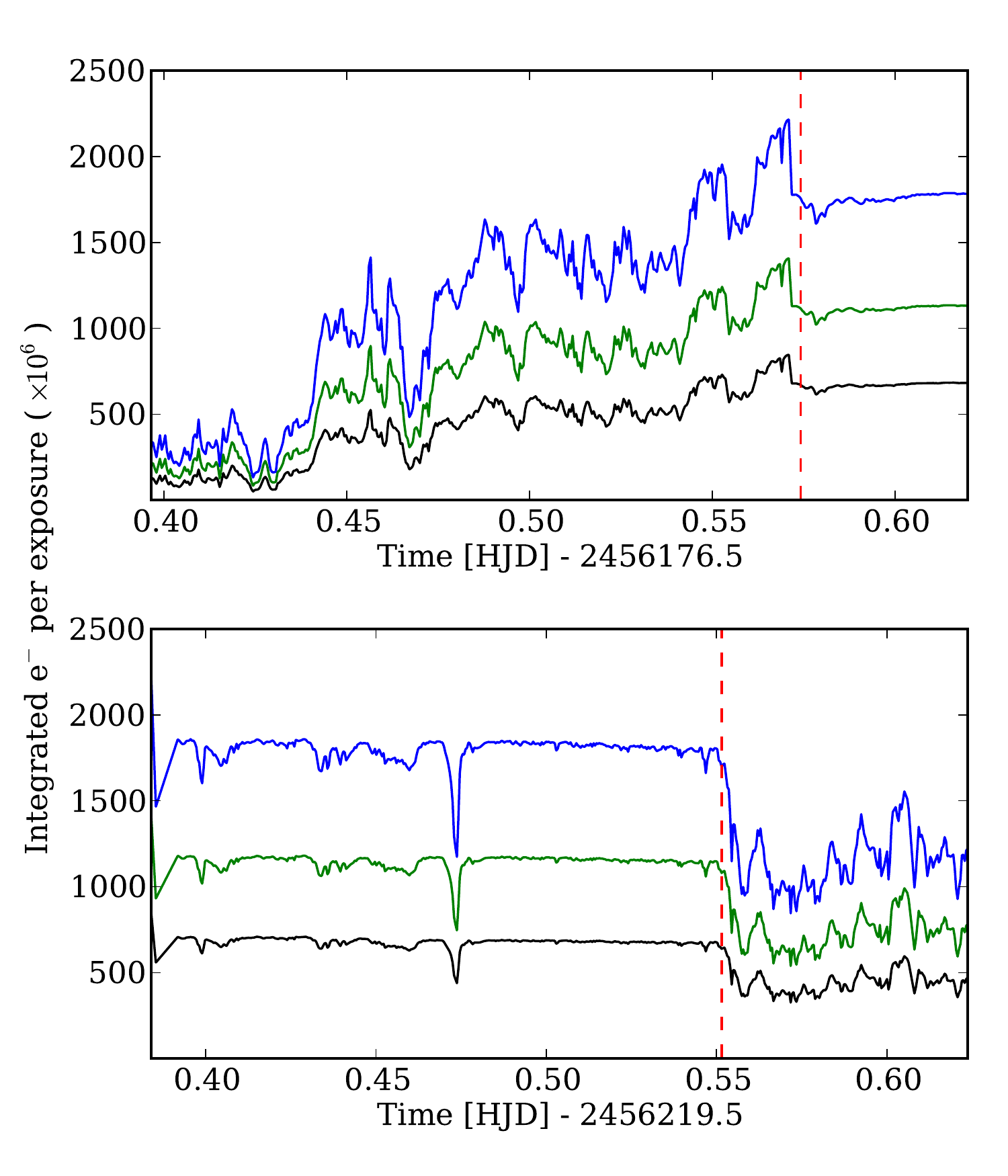}
\includegraphics[width=85mm]{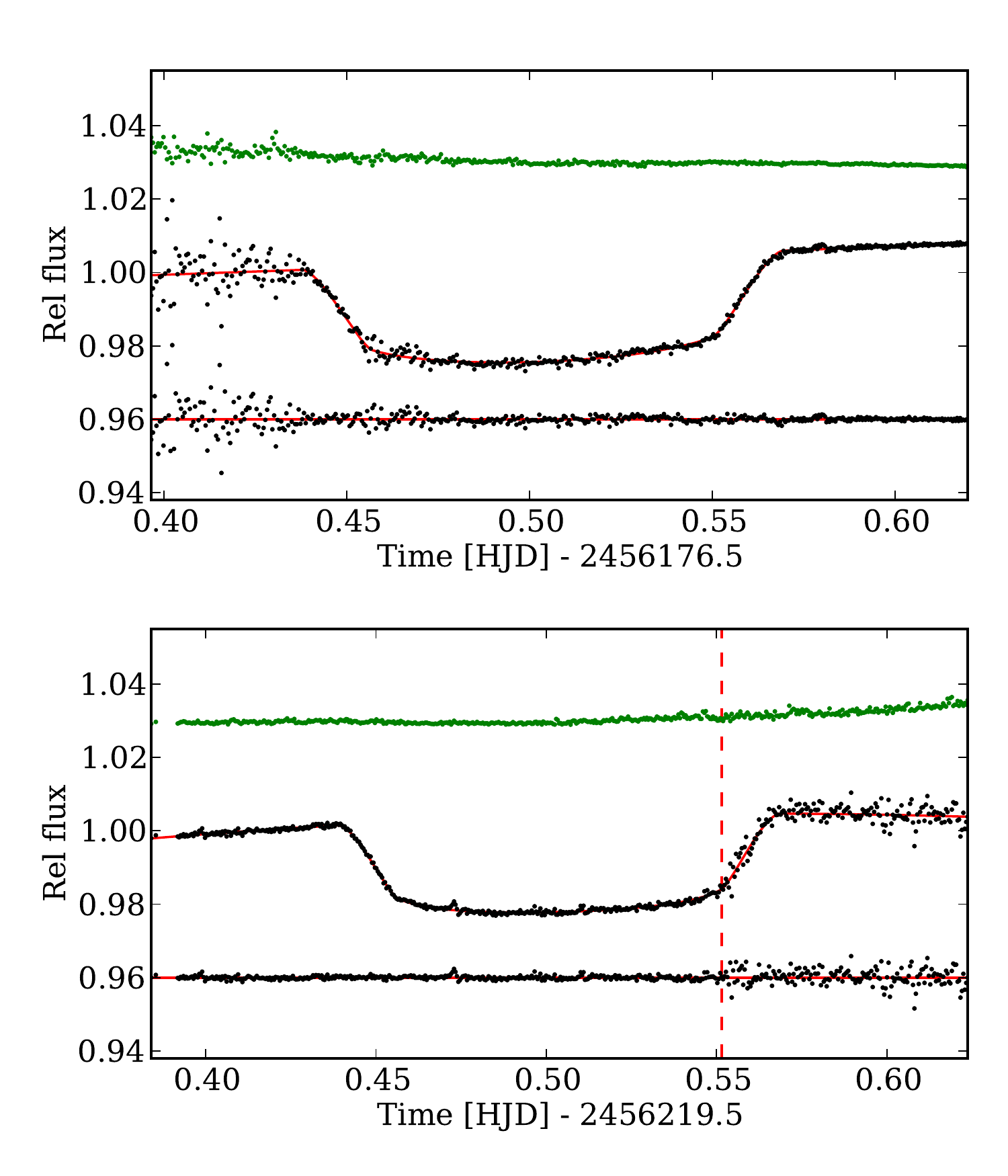}
\caption{Left: Integrated flux over all wavelengths for the target and comparison stars as a function of time, using the same colour scheme as Fig.~\ref{fig:eg_spec}. Right: Differential `white' light curves and the residuals from a best fit model are shown by the black dots (see Sect.~\ref{sect:analysis}), and the relative flux of the two comparison stars are shown by the green dots, offset for clarity. The top and bottom plots represent the first and second visit, respectively. For the first visit, the vertical dashed line indicates the change in exposure time. Overall, the flux gradually increases throughout the observations, resulting in a gradual increase in S/N apparent in the differential light curve. For the second visit, the dashed line represents a sudden change in sky transparency, resulting in a sudden increase in noise in the differential light curve, where the same position is marked.}
\label{fig:raw_lightcurves}
\end{figure*}

\begin{figure*}
\includegraphics[width=180mm]{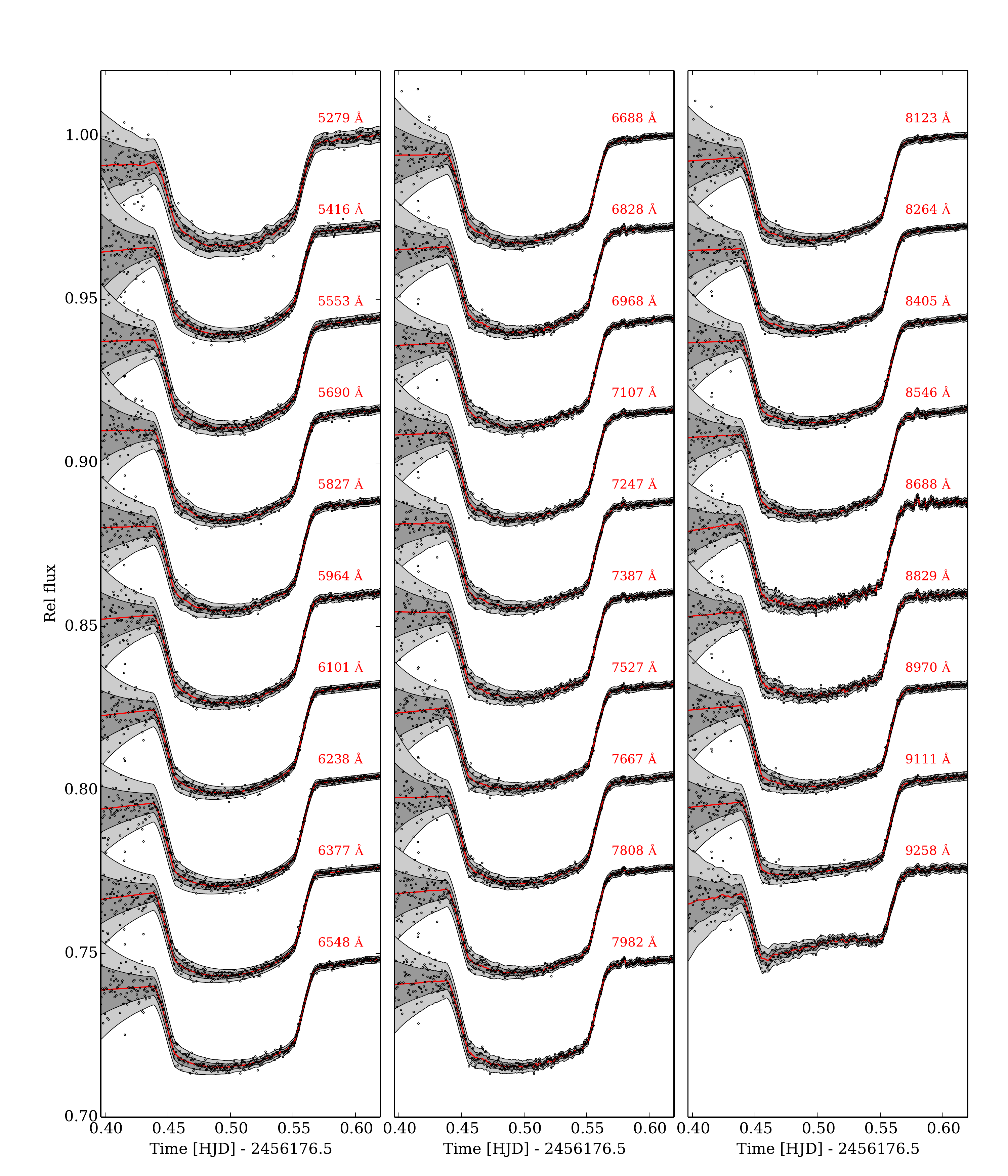}
\caption{Raw spectral light curves for the first transit, offset for clarity. The central wavelengths for each channel are given. The red line indicates the best fit model from the Gaussian process fit to the transit and systematics simultaneously, and the grey shading marks the 1 and 2\,$\sigma$ limits of the best fit GP model.}
\label{fig:rawspec_lightcurves_1}
\end{figure*}

\begin{figure*}
\includegraphics[width=180mm]{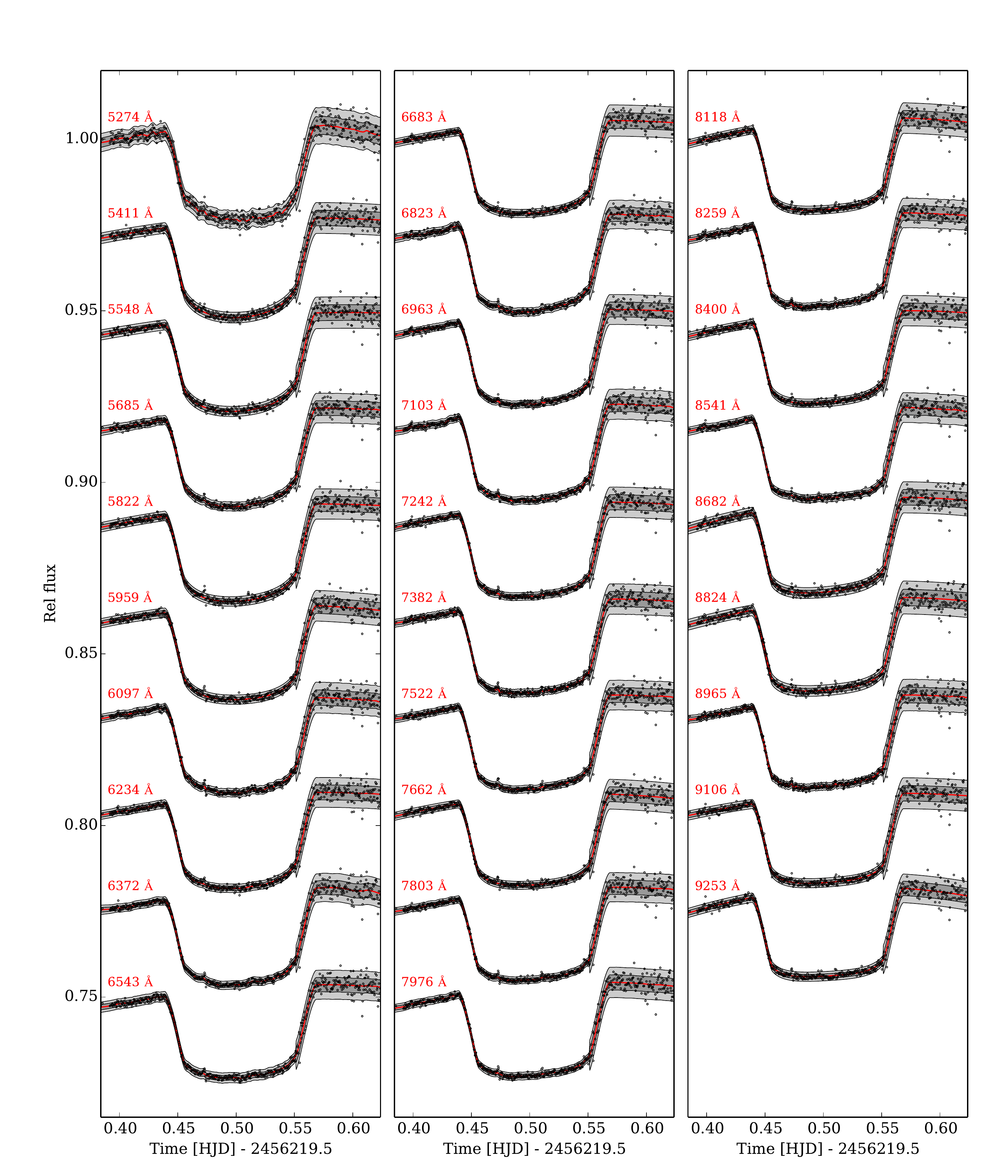}
\caption{Raw spectral light curves for the second transit, offset for clarity. Shading and models are the same as Fig.\ref{fig:rawspec_lightcurves_1}.}
\label{fig:rawspec_lightcurves_2}
\end{figure*}

Examples of extracted spectra of HAT-P-32 and the two comparison stars are shown in Fig.~\ref{fig:eg_spec}. We divided the target spectrum into 29 wavelength bins (each of the three chips were divided into 10 evenly spaced bins, and the faintest channel at the blue end was excluded by the pipeline). Differential `spectral' light curves were produced by dividing the flux of the target star by the combined flux of the comparison stars in each wavelength bin. We also produced a `white' light curve by first integrating over all wavelengths. The raw white flux time-series of the target and comparison stars are shown in Fig.~\ref{fig:raw_lightcurves}, as well as the differential white light curves for the two transits and the differential light curves for the two comparison stars.
The raw fluxes illustrate the variability of the conditions as a function of time, and the effect on the differential light curves is apparent. For the first transit we see a gradual rise in the counts as well as some short term variations, corresponding to a gradual reduction of the white noise in the differential light curve. For the second transit the conditions are stable until egress where the flux suddenly drops and becomes unstable due to cloudy conditions. This corresponds to a sudden increase in the white noise at egress, marked by the vertical dashed line. This motivates our light curve noise models, discussed in the following section. Figs.~\ref{fig:rawspec_lightcurves_1} and \ref{fig:rawspec_lightcurves_2} show the 29 raw, differential spectral light curves for the first and second transit, respectively.

\section{Light Curve Analysis}
\label{sect:analysis}

\subsection{White Light Curve Analysis}
\label{sect:wlc_analysis}

\subsubsection{White noise model}
\label{sect:analysis_wn}

We analysed the light curves in a similar way to \citet{Gibson_2013}. Our light curve model is calculated using the equations of \citet{mandel_agol_2002}, using quadratic limb darkening. We fitted for the central transit time ($T_\text{C}$), the system scale ($a/R_\star$), the planet-to-star radius ratio ($\rho=R_\text{p}/R_\star$), the impact parameter ($b$), the two quadratic limb darkening parameters ($c_1$, $c_2$), and three parameters describing a quadratic function of time for the baseline ($f_\text{oot}$, $T_\text{grad}$, $T_\text{grad,2}$) that we justify at the end of this section. We fixed the period at the value given in \citet{Hartman_2011} of 2.150008 days. Hereafter, the transit model is denoted as $T(\bmath t,\bphi)$, where $\bmath t$ is a vector of time and $\bphi$ is the vector of transit parameters. We denote the vector of flux measurements as $\bmath f$.

Given the variations in the observing conditions for both nights we cannot assume that the white noise is stationary (\ie constant with respect to time). The calculated photon noise does not provide a realistic model of the noise (even with linear rescaling). We therefore opted to construct parametric models of the white noise for each transit; whilst not ideal, we do not expect the somewhat arbitrary choice to affect our results, due to a common mode correction that we discuss later. For the first transit, we found that an exponential decay function provided a good estimate of the noise properties:
\[
\sigma_1(t_n) =  \sigma_a \exp\left(-\frac{n}{\sigma_b}\right)+ \sigma_c,
\]
and for the second transit, a simple, piecewise function was sufficient:
\[
\sigma_2(t_n) = \left\{
\begin{array}{ll}
\sigma_d, & ~\mathrm{if}~n < 380 \\
\sigma_e, & ~\mathrm{if}~n \geq 380 \\
\end{array}
\right.,
\]
where $\{\sigma_a,\sigma_b,\sigma_c,\sigma_d,\sigma_e\}$ are the parameters of the white noise models, and $n$ indexes the time from $0$. These where in part motivated by the raw flux counts plotted in Fig.~\ref{fig:raw_lightcurves}, where we see a gradual increase in the flux for the first transit, and a sudden drop in flux for the second transit, corresponding to a sudden change of observing conditions.

We first assume that the time-varying noise followed these simple functions, but with no correlations between flux measurements. Therefore our likelihood function for each transit is:
\[
p(\bmath f| \bmath t,\bphi) = \mathcal{N} \left (T(\bmath t,\bphi) , \sigma_L(\bmath t)^2\mathbfss{I} \right),
\]
where $\mathcal{N}(\bmu,\mathbf\Sigma)$ is the multivariate normal distribution with mean $\bmu$ and covariance matrix $\mathbf\Sigma$, $ \sigma_L(\bmath t)$ are the white noise models giving the uncertainty of each data point where $L = \{1,2\}$ and \mathbfss{I} is the identity matrix. This likelihood defines a diagonal covariance matrix with each element on the diagonal equal to $\sigma_L(t_n)^2$.

We then fitted both transits simultaneously, by multiplying the two likelihood functions and any prior distributions together to produce the joint posterior probability distribution. A Markov-Chain Monte-Carlo (MCMC) algorithm was used to sample the posterior distribution and obtain marginalised probability distributions for each of our model parameters \citep[see][and references therein for  details on our implementation]{Gibson_2013}. We do not explicitly state priors for all of the parameters, implying uniform, improper priors, and we restrict the limb darkening parameters, $a/R_\star$, $\rho$ and $b$, to be positive using an improper prior of the form:
\[
p(x) = \left\{
\begin{array}{l}
0,~\mathrm{if}~x < 0 \\
1,~\mathrm{if}~x \geq 0 \\
\end{array}
\right..
\]
We also restrict the sum of the limb darkening parameters to be less than 1 using the joint prior:
\[
p(c1,c2) = \left\{
\begin{array}{l}
0,~\mathrm{if}~c_1+c_2 > 1 \\
1,~\mathrm{if}~c_1+c_2 \leq 1 \\
\end{array}
\right..
\]
Four MCMC chains of length 200\,000 were computed, and the first 20\% of each chain was discarded. The light curves and their best fit models are shown in Fig.~\ref{fig:raw_lightcurves}.

Our choice of a 2nd order polynomial of time as the baseline function was justified by computing the Bayes factor for models with a linear and quadratic function of time, which compares the probability of the data under each model (assuming equal prior probabilities for each model) and naturally takes into account model complexity. We first calculate the Bayesian evidence for each model using an importance sampler \citep[see \eg][for details]{Bishop}. Importance sampling uses a proposal distribution from which it is straightforward to draw samples, and evaluates the posterior distribution at each sample point. Expectations are then estimated by weighting each sample using the proposal distribution, thus correcting for any bias due to sampling from the wrong distribution, \ie the expectation of a function $f$ of the probability distribution is given by
\[
\mathbb{E}[f] = \frac{1}{S}\sum_{s=1}^S \frac{p(\bphi_n)}{q(\bphi_s)} f(\bphi_s),
\]
where $\bphi_s$ is the sample point, $p(\bphi_s)$ and $q(\bphi_s)$ are the values of the posterior and proposal distributions at $\bphi_s$, and $S$ is the number of samples. The success of an importance sampler depends on how closely the proposal distribution matches the true posterior. Here, we assume a Gaussian proposal distribution and estimate its mean and covariance from the MCMC distribution, and double the covariance to ensure the posterior is well sampled in case of significant departures from a Gaussian distribution. We take 50\,000 samples from the proposal, and evaluate the posterior of our model at each sample point\footnote{In practice we use the log posterior and calculate the log evidence}. The Bayesian evidence ($E$) is simply the integral under the posterior and is calculated as:
\[
E = \frac{1}{S}\sum_{s=1}^S \frac{p(\bphi_s)}{q(\bphi_s)}.
\]
 We verify our calculation by ensuring that the mean and standard deviation estimated for each model parameter are consistent with those estimated using the MCMC, and by re-running the importance sampler to ensure the evidence calculation was consistent. The Bayes factor is then simply the ratio of the Bayesian evidence for each model. We find that the Bayes factor strongly favours the quadratic model (by many orders of magnitude), and we adopt this as our model of choice. Given how strongly favoured the quadratic model is over the linear model, we do not consider choice of improper priors for the model parameters important.

To further justify our model choice we used the Bayesian Information Criterion (BIC), which uses the maximum likelihood estimate and adds a complexity penalty for each degree of freedom in our model \citep{Schwarz_1978}, and is frequently used in the exoplanet literature \citep[\eg][]{Gibson_2010b,Sing_2011,Crossfield_2013}. Again, this strongly favours a 2nd order polynomial in time. This is not surprising, given that the 2nd order trend can be seen by eye, particularly in the second light curve. We do not consider models where we allow a second order trend in one light curve but not the other, as this could result in more strongly weighting one transit over the other if one model is signficnatly more flexible. We also checked for correlations between the baseline function and the airmass, but found no obvious correlations, indicating that the trend is not a simple function of airmass. This is supported by finding a consistent trend in the spectroscopic light curves, as we would expect this to be less pronounced over narrow wavelength regions if it is the result of colour variations between the target and comparison stars (see Sect.~\ref{sect:specanalysis} and Figs.~\ref{fig:spec_lightcurves_1} and \ref{fig:spec_lightcurves_2}).

After our white noise model fits, correlated noise is clearly present in the light curve residuals, and must be accounted for in the fitting procedure. Given the non-stationary nature of the white noise, we are unable to use all of the methods employed in \citet{Gibson_2013}, as the time-averaging method \citep{Pont_2006} and wavelet method \citep{Carter_2009} are difficult to apply and interpret for variable white noise.

\subsubsection{Gaussian process model}
\label{sect:analysis_gp}

To fit for time-correlated noise simultaneously with our time-dependent white noise model, we employ the flexibility of the Gaussian process (GP) model introduced in \citet{Gibson_2012}, and further applied in \citet{Gibson_2012b}. Application to Gemini/GMOS light curves is also discussed in \citet{Gibson_2013}. We apply similar methods here and refer the reader to these papers for details.

A GP is a collection of random variables, any finite subset of which have a joint Gaussian distribution. In practice, the only difference from our white noise model is that we allow for a non-diagonal covariance matrix with the elements set via a kernel function\footnote{This means the probability distribution is defined for all possible inputs, rather than just for the specific times of our observations. It is this property that allows GPs to make predictions for any arbitrary values of inputs, and so defines a GP as a probability distribution over functions, distinguishing it from a simple multivariate Gaussian distribution.}.
It is fully specified by its mean and covariance:
\[
\label{eq:gp}
p(\bmath f| \bmath t,\bphi,\btheta) = \mathcal{N} \left (T(\bmath t,\bphi) , \mathbf{\Sigma} (\bmath t,\btheta) \right).
\]
Here, $T(\bmath t,\bphi)$ is the transit function (or mean function), and $\mathbf{\Sigma} (\bmath t,\btheta)$ is the kernel (or covariance function) with parameters $\btheta$:
\[
\mathbf\Sigma_{nm} = k({t}_n,{t}_m | \btheta).
\]
As in \cite{Gibson_2013} we use the Mat\'ern 3/2 kernel function to model time-correlations in the data, given by:
\[
k({t}_n, {t}_m | \btheta) = \xi^2 \left( 1+{\sqrt{3}\eta\Delta t} \right) \exp \left( -{\sqrt{3}\eta\Delta t}\right) + \delta_{nm}\sigma_L(t_n)^2,
\]
where $\xi$ is a hyperparameter that specifies the maximum covariance, $\Delta t = |t_n-t_m|$ is the time difference, $\eta$ is the inverse characteristic length scale, and $\delta$ is the Kronecker delta. 
Again, we add the white noise term, $\sigma_L(t_n)^2$, to the diagonal of the covariance matrix to account for the non-stationary white noise. Whilst the GP could in principle model the quadratic baseline discussed in the previous section, we choose to keep this as part of the mean function as it likely has a separate origin than the higher frequency correlated noise remaining in the white noise residuals, and it is clearly supported by the data.

We defined a hyperprior for the inverse length scale, $\eta$. This took the form of a Gamma distribution with shape parameter unity, given by
\[
p(\eta) = \left\{
\begin{array}{ll}
0, & ~\mathrm{if}~\eta < 0 \\
\dfrac{1}{l} \exp\left(-\eta/l_\eta\right), & ~\mathrm{if}~\eta \geq 0 \\
\end{array}
\right.,
\]
where $l_\eta$ is the length scale of the hyperprior. We set the length scale to $200$ as in \citet{Gibson_2013}. For $\xi$, instead of applying a Gamma prior, we fitted for $\log\xi$. This is equivalent to placing a prior on $\xi$:
\[
p(\xi) = \frac{1}{\xi}
\]
for positive $\xi$, and is a natural choice for a scale parameter. These priors were not intended to influence the results of the inference, but rather to ease convergence of the MCMC chains when the parameters are poorly constrained, and to discourage the GP model from fitting very high frequency systematics. Indeed, when the parameters are strongly constrained by the data, the likelihood should dominate the posterior distribution, and the (hyper-)priors should have negligible influence.

\begin{figure}
\includegraphics[width=85mm]{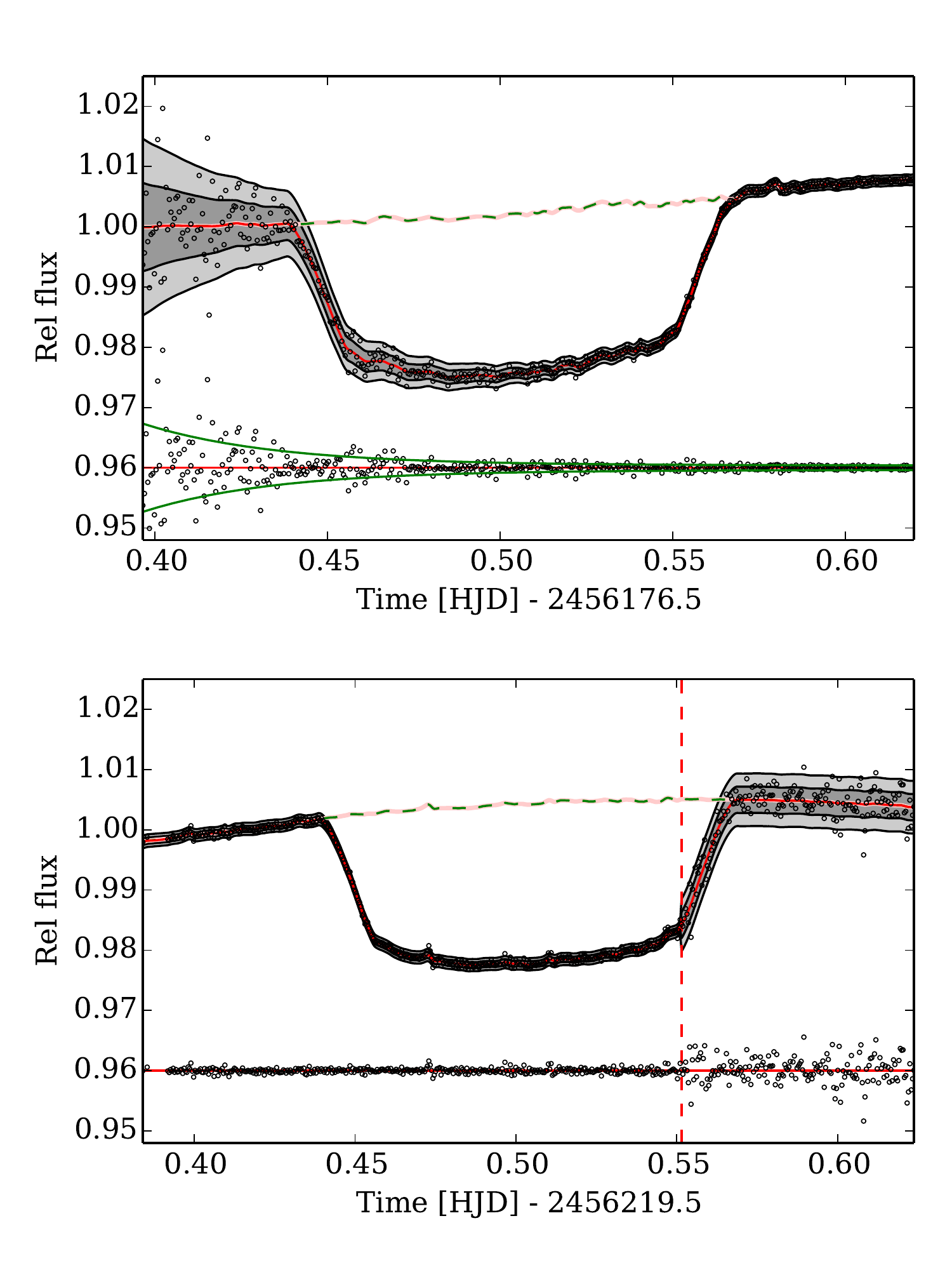}
\caption{White light curves for the two transits, fitted with a Gaussian process model, along with their residuals from the best fit model, offset for clarity. The solid line is the best fit model, and the dashed green line is the systematics model without the transit function (but with the normalisation function). The grey regions represent the $1\sigma$ and $2\sigma$ predictions of the GP model. The solid green line for the residuals of the first transit is the best fit white noise model, and the vertical dashed line for the second transit marks the changepoint in the white noise model.
}
\label{fig:GP_lightcurves}
\end{figure}

\begin{figure}
\includegraphics[width=85mm]{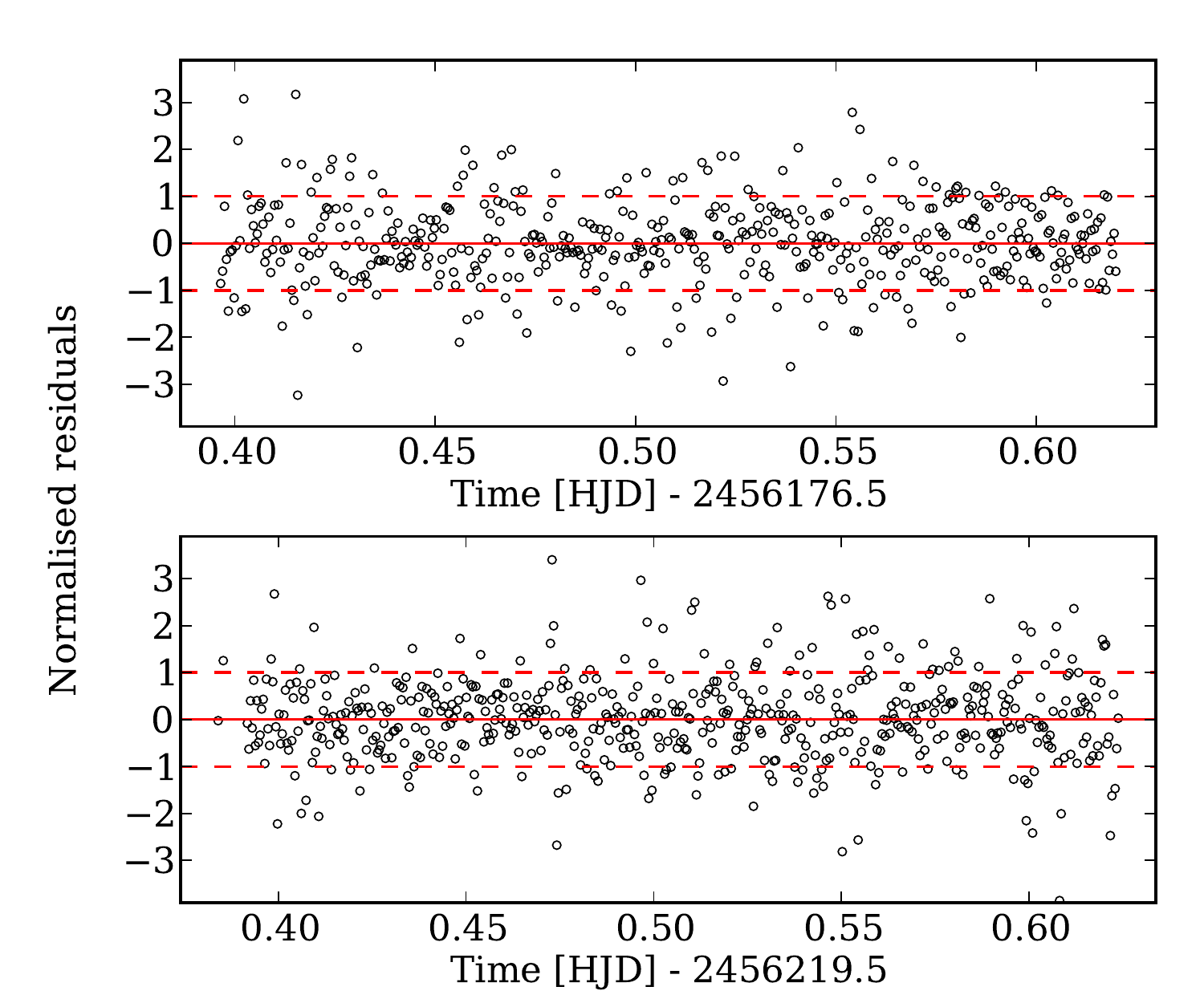}
\caption{Normalised residuals from the GP models, \ie the residuals from the best fit model divided through by their best fit uncertainties. These should be distributed around zero with a variance of 1 if the noise model aptly characterises the noise.}
\label{fig:norm_res}
\end{figure}

Inference is performed in the same way as for the white noise model. In this context hyperparameters are treated in exactly the same way as parameters of our transit model.
The likelihoods and priors are combined, and we sample from the joint posterior probability distribution using MCMC to fit both transits along with their systematics models simultaneously. Note that each transit has separate values (and hyperpriors) for $\xi$ and $\eta$.
The white light curves along with their best fit models are shown in Fig.~\ref{fig:GP_lightcurves}, as well as the best fit systematics models projected without the transit function. The GP captures the time-correlated systematics as well as the non-stationary white noise.
Fig.~\ref{fig:norm_res} shows the normalised residuals after fitting our GP model, in other words the residuals after dividing through by our inferred error function, $\sigma_L(\bmath t)$. These should be (and are) distributed with zero mean and variance of 1, thus validating our choice of error function. 

Our GP model has significant advantages over parametric models, in that we can avoid having to specify an unknown function in closed form to describe the systematics, but rather place a probability distribution over a broad and complex class of plausible functions with only a few parameters. We then marginalise out our ignorance of these functions, analogous to marginalising out model parameters in `normal' Bayesian inference. Furthermore, a GP is intrinsically Bayesian, and finds the simplest model that explains our data, automatically invoking Occam's Razor. Our choice of kernel is justified in \citet{Gibson_2013}, but we emphasise that any sensible choice of kernel is superior to assuming a parametric model.

\subsection{Spectral light curve analysis}
\label{sect:specanalysis}

Our raw, differential spectral light curves consist of 29 wavelength channels for each of the two transits. These are shown in Figs.~\ref{fig:rawspec_lightcurves_1} and \ref{fig:rawspec_lightcurves_2}. We proceed by fitting the light curves with the same GP model described for the white light curves. The system parameters $T_\text{C}$, $a/R_\star$ and $b$ are now fixed at their best fit values, as we are interested in the conditional distribution of $\rho$ for each wavelength channel to obtain our transmission spectrum; therefore anything that should be the same for each wavelength channel is fixed. This leaves $\rho$, the limb darkening parameters, the normalisation parameters, and the kernel hyperparameters (\ie systematics parameters) to freely vary for each light curve fit. We fitted the wavelength channels for the two transits independently. The best fit models are shown in Figs.~\ref{fig:rawspec_lightcurves_1} and \ref{fig:rawspec_lightcurves_2}, along with the best-fit GP noise models.

After fitting the wavelength channels and inspecting the residuals, it was clear that the spectral light curves for each transit showed common mode systematics. The same effect was seen for GMOS-South data in \cite{Gibson_2013}, and we corrected for it in the same way. Each spectral light curve was divided through by the systematics model, which consisted of the normalisation function plus the systematics model from the GP; in other words the dashed lines in Fig.~\ref{fig:GP_lightcurves}. The systematics are clearly a multiplicative (sensitivity) effect, as the common mode systematics have the same amplitude in the {\it normalised} light curves, despite each channel having a large variation in raw flux. We note that the GP systematics function was additive rather than multiplicative, but this effect is negligible providing the GP does not model large amplitude, long term trends, and is why we preferred keeping the 2nd order polynomial as the baseline function, rather than letting the GP account for this.
Prior to fitting, we normalised by a constant value from the median of the out-of-transit flux.

Further to this common mode correction, we noticed that the residuals from the spectral light curves were correlated with residuals from other wavelength channels. We therefore implement another common mode correction, where we subtract the residuals from the corresponding white light curve fits from each of the wavelength channels. This was implemented after dividing through by the normalisation function plus systematics model, and after setting the out-of-transit flux to 1. The origin of this effect is probably related to spatial variation in the atmospheric throughput due to clouds that cannot be accounted for by the comparison stars. Applying such common mode corrections is valid when extracting transmission spectra, as we are only interested in the differential transit depths. Indeed, we are conditioning on a common systematics model just as we condition on common physical parameters.

These corrections enabled us to reach precision of $\approx$1.4--2.5 times the calculated Poisson noise, where for each spectral light curve, we take the median value of the ratio between the fitted white noise model and calculated photon limit. The subtraction of the residuals typically improved the (median) noise estimate by a factor of $\approx$1.6--1.8.
For the clean sections of the white light curves, we reach precision of $\approx$10 times the photon limit, and considerably higher for the noisier parts of the white light curves (as much as 40--50 times). This clearly shows there is a significant non-white noise component in the differential white light curve, which is even more important when observing during variable weather conditions, and further justifies our use of a common mode correction. Note that this does not take into account any remaining correlated noise in the light curves.

Finally, we noticed that some of the light curve fits were skewed by outliers in the spectral light curves. We decided to cull outlying points more that 4$\sigma$ from the predictive GP distribution from each individual fit, and repeated our fits. In practice this resulted in a few points ($\leq4$) being removed from each light curve.

\begin{figure*}
\includegraphics[width=180mm]{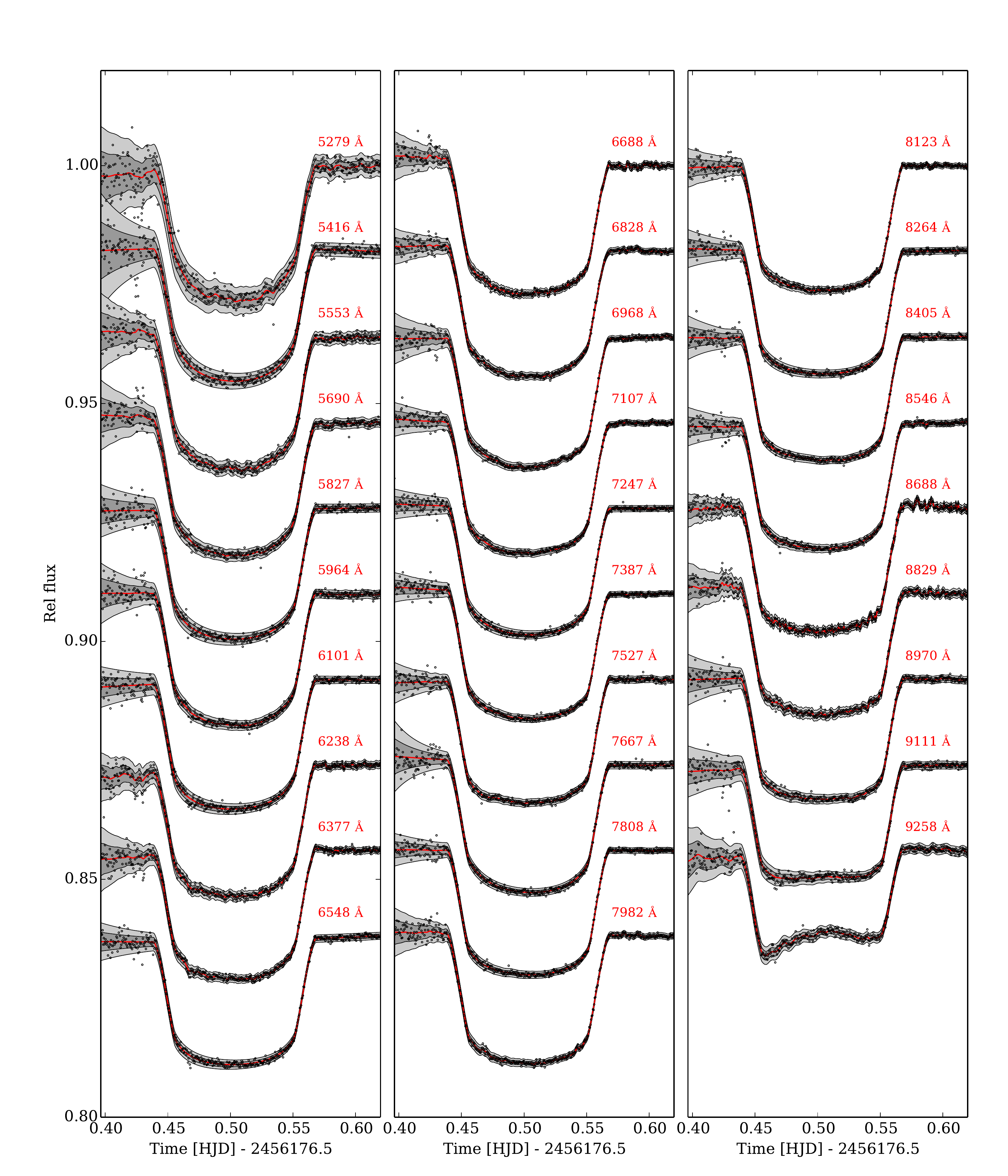}
\caption{Spectral light curves for the first transit, after common mode corrections are applied, and offset for clarity. The central wavelengths for each channel are given. The red line indicates the best fit model from the Gaussian process fit to the transit and systematics simultaneously, and the grey shading marks the 1 and 2\,$\sigma$ limits of the best fit GP model.}
\label{fig:spec_lightcurves_1}
\end{figure*}

\begin{figure*}
\includegraphics[width=180mm]{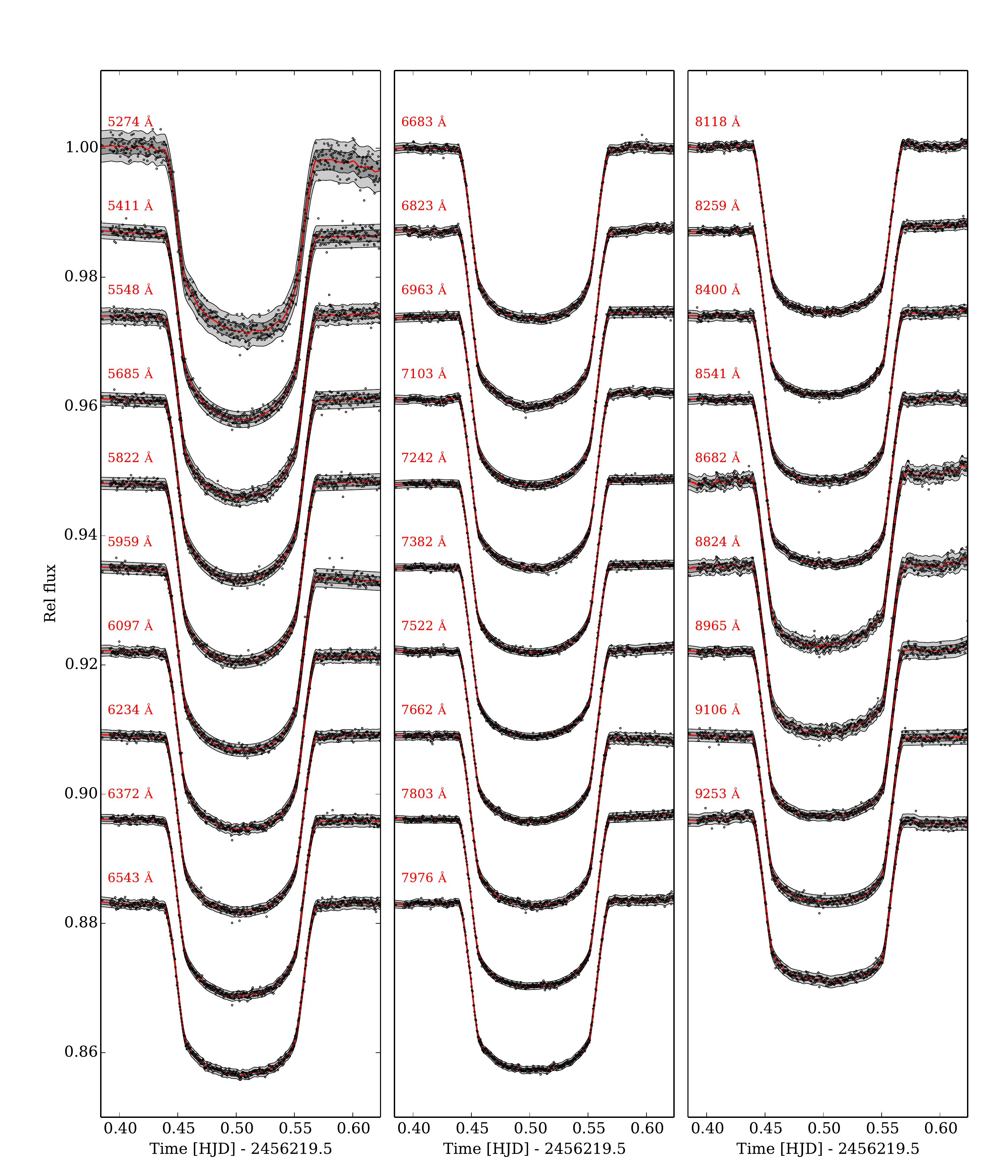}
\caption{Spectral light curves for the second transit, after common mode corrections are applied, and offset for clarity. Shading and models are the same as Fig.\ref{fig:spec_lightcurves_1}. The offsets between the light curves have been reduced for greater clarity.}
\label{fig:spec_lightcurves_2}
\end{figure*}

The light curves after implementing these common mode corrections are shown in Figs.~\ref{fig:spec_lightcurves_1} and \ref{fig:spec_lightcurves_2}. Most of the large scale systematics are removed, and also variation in noise due to the variable conditions is considerably reduced. For completeness, we still fit all light curves using the same noise models, as the exponential decay is still present in the first transit. Whilst the sudden change in noise properties is still present to some degree in the second transit, the noise is much closer to stationary, and our model is therefore much less dependent on the assumed noise model. Furthermore, not all of the time-correlated noise is common mode, and we employ the same GP model to account for any remaining systematics in our fits. For all fits, we ran four MCMC chains of length 100\,000 and checked for convergence as before. As in \citet{Gibson_2013}, for some runs not all parameters fully converged, with the GR statistic a few percent form unity. This is mostly related to the relation between $\xi$ and $\eta$ parameters, where $\eta$ is only constrained by the prior when $\xi$ is close to zero, and also the more complex noise model and baseline used for these data. However, multiple runs have verified that our transmission spectra are not significantly altered, and this does not effect our final results.

For convenience, we now summarise the steps taken to produce our final transmission spectra:
\begin{itemize}
\item The spectral time series for the target and comparison stars are extracted from the raw data, and used to produce differential light curves integrated over all wavelengths (the `white' light curves), and for 29 spectral bins from $520-930$\nm\ and $\approx14$\nm\ wide.
\item The white light curves for both visits are fitted simultaneously to derive the physical parameters of the HAT-P-32 system and the Gaussian process noise model, which includes correlated noise and time-varying white noise.
\item To account for common mode systematics, each spectral channel is divided through by the best-fitting systematics model from the corresponding white light curve. The systematics models are shown by the dashed lines in Fig.~\ref{fig:GP_lightcurves}.
\item After normalisation, the residuals from the corresponding white light curve are subtracted from each spectral channel, to account for correlated noise between the channels. The effects of the last two steps are to change the light curves from Figs.~\ref{fig:rawspec_lightcurves_1} and \ref{fig:rawspec_lightcurves_2} to Figs.~\ref{fig:spec_lightcurves_1} and \ref{fig:spec_lightcurves_2}.
\item To produce our transmission spectra, we extract the posterior distribution of the planet-to-star radius ratio for each wavelength channel with the same GP model used for the corresponding white light curves. This is conditioned on the common physical parameters and common mode systematics, but marginalised over individual systematics models.
\end{itemize}

\section{Results and Conclusions}
\label{sect:results}

\subsection{White light curve analysis}

Results for the simultaneous GP fit to the white light curves are shown in Tab.~\ref{tab:wlc_results}. The fitted system parameters are consistent with those reported in \citet{Hartman_2011}. In general, our uncertainties are slightly larger, given the noisy ingress and egress for the first and second transits, respectively, and that we were fitting for a systematics model and limb darkening parameters. Our fitted limb darkening parameters are broadly consistent with the $i$ band values reported in \citet{Hartman_2011}, and the $R$ and $I$ filters given in the tables of \citet{Claret_2004} using stellar parameters from the discovery paper, although we do not attempt a detailed calculation for the effective passband or the spectroscopic light curves. We note that the transit depth (and therefore planet radius) reported in \citet{Hartman_2011} is perhaps diluted by the contaminant star. This will dilute the transit depth by a factor of ($1+q$), where $q$ is the ratio of the flux from the target and contaminant star in a particular passband. The transit parameters were inferred from $i$, $z$ and $g$ bands, and we caution that the HAT-P-32b's radius could be marginally larger than that reported in \citet{Hartman_2011}, depending on the apertures used for the photometry and the relative contribution to the final parameters of each of the light curves.

Using the distributions from the MCMC chains, we calculate further system parameters for HAT-P-32, also reported in Tab.~\ref{tab:wlc_results}. Where the distributions were not available, we generated draws from normal distributions from the values reported in the literature and combined them with those inferred from the MCMC chains. This included the stellar mass and radius ($M_\star$ and $R_\star$) and the planetary mass ($M_p$). We calculated values for the transit duration ($T_{14}$),  inclination ($i$), planet radius ($R_p$), planet density ($\rho_p$), log surface gravity ($\log g_p$), and the equilibrium temperature ($T_p$). Again, these distributions are consistent with the results of \citet{Hartman_2011}.

\begin{table}
\caption{Fitted parameters for HAT-P-32b from the white light curves, given for the fully marginalised Mat\'ern 3/2 Gaussian process model, as well as parameters derived from the posterior distribution.}
\label{tab:wlc_results}
\begin{tabular}{lll}
\hline
\noalign{\smallskip}
Parameter & Value & Unit\\
\hline
\noalign{\smallskip}
\multicolumn{3}{l}{Global parameters:} \\\noalign{\smallskip}
~$P^\alpha$ & $2.150008$ (fixed$$) & - \\\noalign{\smallskip}
~$a/R_\star$ & $6.091^{+0.036}_{-0.047}$ & - \\\noalign{\smallskip}
~$R_\text{p}/R_\star$ & $0.1515^{+0.0012}_{-0.0012}$ & - \\\noalign{\smallskip}
~b & $0.093^{+0.071}_{-0.065}$ & - \\\noalign{\smallskip}
~$c_1$ & $0.279^{+0.070}_{-0.074}$ & - \\\noalign{\smallskip}
~$c_2$ & $0.254^{+0.123}_{-0.122}$ & - \\\noalign{\smallskip}
\multicolumn{3}{l}{Transit 1:} \\\noalign{\smallskip}
~$T_\text{C}$ & $2456177.0031611^{+0.0002472}_{-0.0002410}$ & HJD$_{UTC}$ \\\noalign{\smallskip}
~$f_{oot}$ & $1.0023^{+0.0007}_{-0.0006}$ & - \\\noalign{\smallskip}
~$T_{grad}$ & $0.0014^{+0.0002}_{-0.0002}$ & - \\\noalign{\smallskip}
~$T_{grad,2}$ & $0.00023^{+0.00013}_{-0.00018}$ & - \\\noalign{\smallskip}
~$\log\xi$ & $-3.19^{+0.26}_{-0.15}$ & - \\\noalign{\smallskip}
~$\eta$ & $65.04^{+31.49}_{-30.24}$ & - \\\noalign{\smallskip}
~$\sigma_a$ & $0.00691^{+0.00083}_{-0.00069}$ & - \\\noalign{\smallskip}
~$\sigma_b$ & $78.4^{+8.3}_{-7.6}$ & - \\\noalign{\smallskip}
~$\sigma_c$ & $0.00038^{+0.00004}_{-0.00005}$ & - \\\noalign{\smallskip}
\multicolumn{3}{l}{Transit 2:} \\\noalign{\smallskip}
~$T_\text{C}$ & $2456220.0036343^{+0.0001864}_{-0.0001848}$ & HJD$_{UTC}$ \\\noalign{\smallskip}
~$f_{oot}$ & $1.0044^{+0.0006}_{-0.0005}$ & - \\\noalign{\smallskip}
~$T_{grad}$ & $0.0010^{+0.0001}_{-0.0001}$ & - \\\noalign{\smallskip}
~$T_{grad,2}$ & $-0.00043^{+0.00008}_{-0.00010}$ & - \\\noalign{\smallskip}
~$\log\xi$ & $-3.41^{+0.30}_{-0.21}$ & - \\\noalign{\smallskip}
~$\eta$ & $42.27^{+36.55}_{-21.63}$ & - \\\noalign{\smallskip}
~$\sigma_d$ & $0.00048^{+0.00002}_{-0.00002}$ & - \\\noalign{\smallskip}
~$\sigma_e$ & $0.00215^{+0.00012}_{-0.00011}$ & - \\\noalign{\smallskip}
\multicolumn{3}{l}{Ephemeris:} \\\noalign{\smallskip}
~$T_0$ & $2454942.898449\pm0.000077$ & HJD \\\noalign{\smallskip}
~$P$ & $2.1500085\pm0.0000002$ & days \\\noalign{\smallskip}
\multicolumn{3}{l}{Derived parameters:} \\\noalign{\smallskip}
~$T_{14}$ & $0.12959^{+0.00059}_{-0.00057}$ & days \\\noalign{\smallskip}
~$i$ & $89.12^{+0.61}_{-0.68}$ & deg \\\noalign{\smallskip}
~$M_\star^\beta$ & $1.160^{+0.041}_{-0.041}$ & $M_\odot$ \\\noalign{\smallskip}
~$R_\star^\beta$ & $1.219^{+0.016}_{-0.016}$ & $R_\odot$ \\\noalign{\smallskip}
~$M_p^\beta$ & $0.860^{+0.164}_{-0.164}$ & $M_J$ \\\noalign{\smallskip}
~$R_p$ & $1.796^{+0.028}_{-0.027}$ & $R_J$ \\\noalign{\smallskip}
~$\rho_p$ & $0.18^{+0.04}_{-0.04}$ & g cm$^{-3}$ \\\noalign{\smallskip}
~$\log\,g_p$ & $2.82^{+0.08}_{-0.09}$ & [cgs] \\\noalign{\smallskip}
~$T_p$ & $1779^{+26}_{-26}$ & K \\\noalign{\smallskip}
\hline
\noalign{\smallskip}
\multicolumn{3}{l}{{\footnotesize\,$^\alpha$\,Fixed at the value of \citet{Hartman_2011} for model fitting.}} \\
\multicolumn{3}{l}{{\footnotesize\,$^\beta$\,Adopted from \citet{Hartman_2011}.}} \\
\noalign{\smallskip}
\end{tabular}
\end{table}

\subsection{Transit Ephemeris}

We calculated a new ephemeris for HAT-P-32b using our two transit times, and the ephemeris reported in \citet{Hartman_2011}. We fitted a straight line of the form
\[
T_\text{C}(E) = T_\text{C}(0) + PE
\]
to the transit times, where $P$ is the Period and $E$ is the epoch. We selected $E=0$ to correspond to the centre of mass as closely as possible (weighted as ${1}/{\sigma_{T_C}^2}$), resulting in epochs of -243, 574 and 594 for the discovery epoch and our two light curves. In order to account for prior information on the Period (as we use the ephemeris rather than the individual transit times from \citealt{Hartman_2011}), we use a Gaussian prior with a mean and standard deviation as reported in \citet{Hartman_2011}. This has negligible influence on the resulting ephemeris, which is dominated by the long baseline between the discovery observations and our Gemini light curves. The updated ephemeris is reported in Tab.~\ref{tab:wlc_results}.

\subsection{Transmission spectrum}
\label{sect:trans_spectrum}

\begin{figure}
\includegraphics[width=85mm]{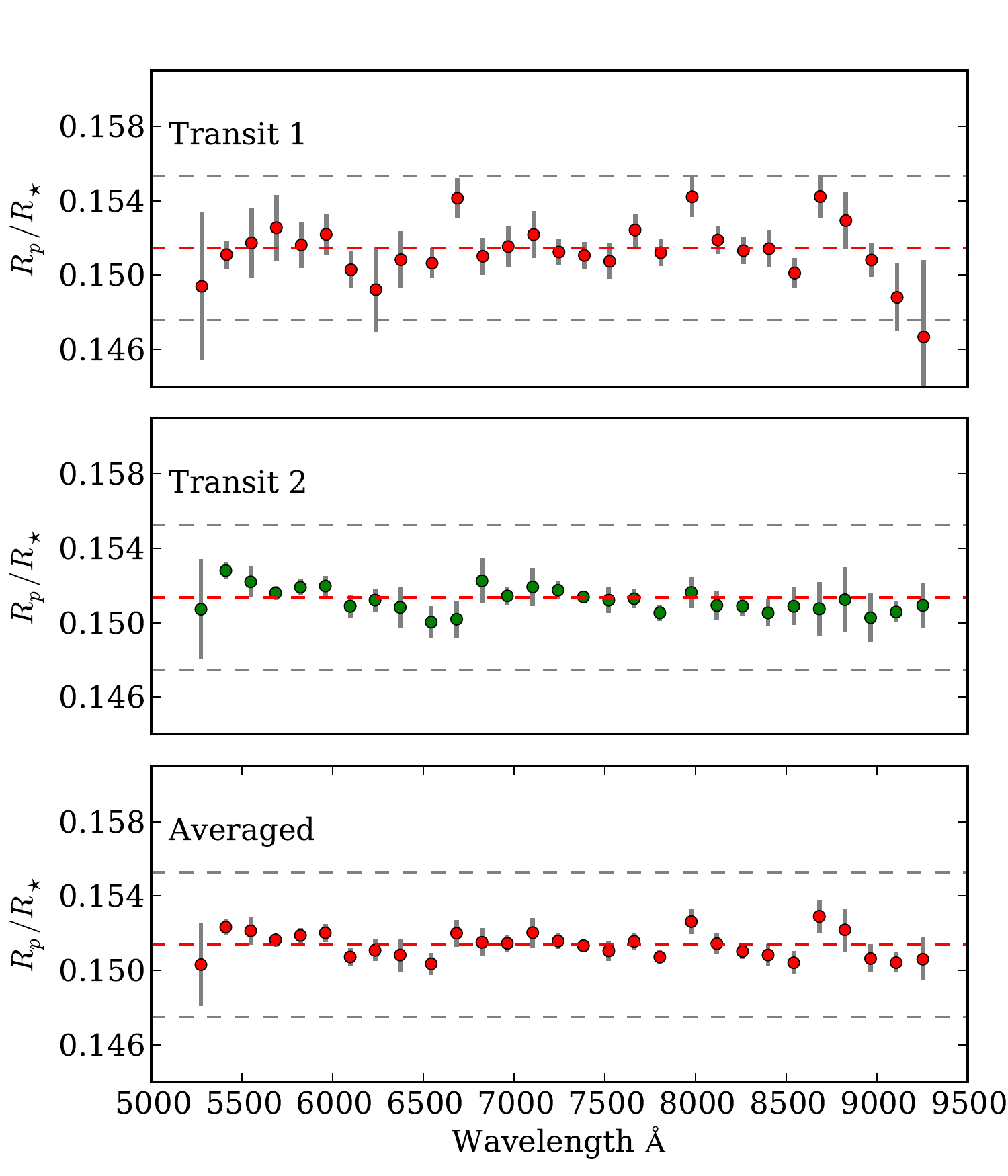}
\caption{Transmission spectra of HAT-P-32b for the first transit, second transit and a weighted average of both. The dashed lines are the weighted mean of each spectra, plus and minus three atmospheric scale heights.}
\label{fig:trans_spec_all}
\end{figure}

The transmission spectra are shown in Fig.~\ref{fig:trans_spec_all} for the first and second transits, and for a combined spectrum derived from a weighted average of the two transits. These results are reported in Tab.~\ref{tab:trans_spec}, along with the fitted limb darkening coefficients. We ignore the small $\sim$5\,\AA\ shifts in the central wavelengths when combining the spectra. The dashed horizontal lines correspond to the weighed average of the spectra, plus and minus three pressure scale heights of the atmosphere, where 1 scale height is $\approx$1100\,km when assuming the equilibrium temperature.
The most striking property of the spectra is how flat they are, and they are all marginally consistent with a flat spectrum, giving reduced $\chi^2$ of 1.23, 1.03 and 1.29 for the two transits and the combined spectrum, respectively.
The slightly larger reduced $\chi^2$ for the combined spectrum perhaps indicates that the two individual spectra are marginally inconsistent. On closer inspection, several deviant points at about 670, 800 and 870\nm\ in the first spectrum are mainly responsible.
The limb darkening coefficients for the two runs are largely consistent, vary quite smoothly from wavelength to wavelength, and are broadly consistent with the white light curve values at the central wavelengths, thus are not responsible for these discrepancies\footnote{In some cases the limb darkening parameters are better constrained than the white light curve fits; this is simply a result of fixing common transit parameters.}.
The disagreement at $\approx$\,670 and 800\nm\ are perhaps related to stars lying at the edges of the GMOS detectors.

To be conservative, given that the first transit has the larger systematics, larger uncertainties and the more convoluted noise model, we choose to focus the rest of our interpretation on the second visit's transmission spectrum, as it is easier to trust the inference using this simpler noise model. We note that this will not significantly change our conclusions. Another GP could in principle be used to construct models for the time-varying white noise and marginalise out our ignorance of its parametric form, but we choose not to proceed with a more complex model here, which would complicate the inference further. We are cautious not to over interpret any small `features' in the spectra such as deviant points in the first transit, or low amplitude ($\lesssim$1 scale height) variations, as these vary between the two visits, and there may still be correlations between neighbouring points in the spectra, given the nature of common mode systematics. However, we can immediately rule out the presence of broad, spectral features larger than about one atmospheric scale height, and it is this constraint from which we infer information about the atmosphere of HAT-P-32b.

\begin{figure*}
\includegraphics[width=140mm]{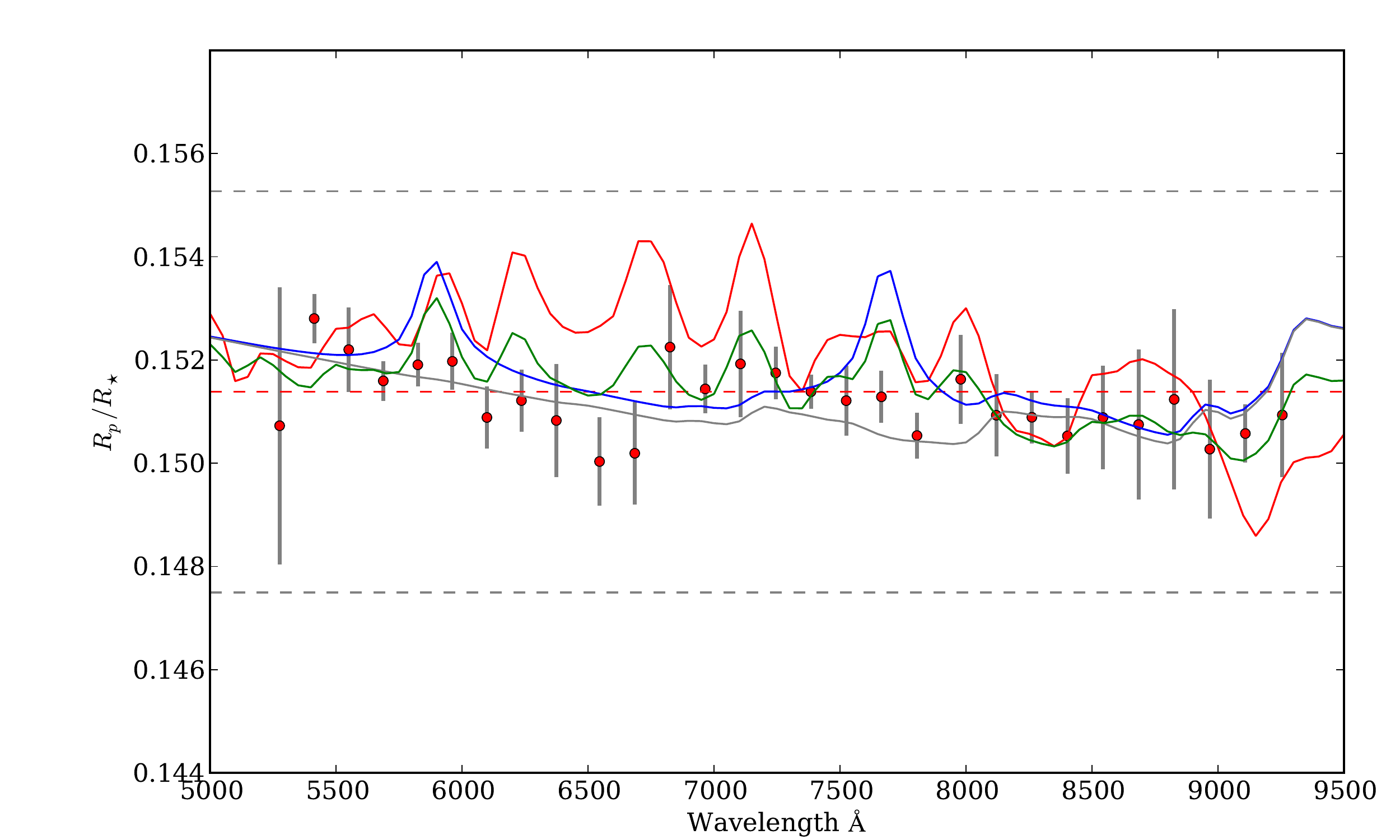}
\caption{Transmission spectrum of HAT-P-32b for the second transit. The dashed horizontal lines represent the weighted mean and plus and minus three scale heights, where one scale height $\approx$1100\,km. The solid lines represent various forward models produced using the {\sc nemesis} code (see Sect.\ref{sect:trans_spectrum}). The red line is a model with solar abundances of Na, K, TiO, VO and H$_2$O in a H$_2$ dominated atmosphere. The blue is the same without TiO and VO. The green is a Na, K, TiO, VO and H$_2$O model but with substellar ($\sim10^{-3}$ solar) abundances of TiO and VO. Finally the grey model is with only H$_2$O.}
\label{fig:trans_spec}
\end{figure*}

Fig.~\ref{fig:trans_spec} shows the transmission spectrum of the second visit, again with the weighted average and plus and minus three scale heights indicated by the dashed lines. Given the degeneracy involved in extracting information from a flat spectrum, we do not attempt a full retrieval of the atmospheric parameters. Rather, we compare our transmission spectrum to forward models produced using the {\sc nemesis} retrieval tool \citep{Irwin_2008}, a radiative transfer code originally developed to investigate the atmospheres of Solar system planets, and recently adapted for exoplanet emission and transmission spectra \citep[\eg][]{Lee_2012,Barstow_2013a,Barstow_2013b}. These models are plotted alongside the data in Fig.~\ref{fig:trans_spec}, and we consider the effects of Na, K, TiO, VO and H$_2$O in a H$_2$ dominated atmosphere, as these are the main atomic and molecular features we expect to see at this temperature and wavelength range. The red line is a model with solar abundances of Na, K, TiO, VO and H$_2$O. The large features are mainly due to TiO and VO, and clearly, if TiO and VO are present in the atmosphere, they have significantly lower mixing ratios.
The blue line shows the same model now without TiO and VO, leaving large spectral features from Na and K wings that we can confidently rule out with our observations.
However, it is possible that with small amounts of TiO and VO, we can significantly mask the Na and K features, and obtain a spectrum that is reasonably consistent with our data.
This is shown by the green model, which is the same as the red model, only with $10^{-3}$ the abundances of TiO and VO.
Our flat spectrum could also be produced by a clear atmosphere with only H$_2$O features, shown by the grey line, although we consider this explanation unlikely as we expect Na and K to be present in the atmosphere from both models and observations of hot Jupiters \citep[\eg][]{Fortney_2008,Huitson_2012,Sing_2012}, and require significantly lower temperatures for them to condense out of the atmosphere \citep{Burrows_2000}. Another obvious explanation for our data would be the presence of clouds in the upper atmosphere, which act as a grey absorber and mask potential spectral features. The most extreme effect of clouds would be to create a completely flat spectrum, consistent with the dashed red line, although intermediate scenarios are possible, where the clouds dilute some of the spectral features. We do not attempt to create a more detailed physical cloud model.

To supplement our transmission spectra, we also attempted a targeted search for narrow Na and K cores using a similar strategy to previous studies \citep[\eg][]{Charbonneau_2002, Huitson_2012, Sing_2012}. These features originate at low pressures and may be visible above any hypothetical cloud deck or TiO/VO absorption, and have been observed in several hot Jupiters \citep[\eg][]{Huitson_2012, Sing_2012}. Light curves from narrow spectral regions of width $15-60$\,\AA\ centred on the Na and K cores were extracted and fitted in the same way as the broadband spectral light curves, and we applied the same common mode corrections. As the spectral regions narrowed, the resulting increase in the photon noise resulted in the GP systematics models being poorly constrained. Furthermore, the systematics are considerably larger with a spectral width of 15\,\AA. Unsurprisingly, 140\,\AA\ bins centred on the Na and K features resulted in a planet-to-star radius consistent with a flat spectrum and with similar uncertainties to the surrounding data points. Consequently, we were unable to detect or place meaningful constraints on the presence of Na and K cores. 

This leaves us with two plausible models for our data: 1) Clouds in the upper atmosphere, masking atomic and molecular species resulting in a flat spectrum, and 2) A cloud free atmosphere with Na and K, and with TiO and VO at substellar levels ($\sim10^{-3}$ solar), represented by the green line in Fig.~\ref{fig:trans_spec}. Clouds and/or haze are emerging as a significant feature of many hot Jupiter atmospheres. For example, they are required to explain the broad transmission spectrum of HD 189733b \citep{Pont_2008,Sing_2012,Gibson_2012b,Pont_2013}, and probably to account for diluted molecular features in XO-1b and HD 209458b at infrared wavelengths \citep{Deming_2013}. On the other hand, the presence of TiO and VO has been suggested as a possible explanation for temperature inversions in hot Jupiter atmospheres \citep[\eg][]{Fortney_2008}, and it has been detected in the atmosphere of HD 209458b, albeit tentatively \citep{Desert_2008}.
Only small ($\sim10^{-3}$ solar) abundances of TiO are necessary to explain this potential TiO feature; however, we can rule out a HD 209458b-like atmosphere for HAT-P-32b, given the lack of Na wings. Understanding the abundance of TiO/VO in the atmosphere of hot Jupiters is a complex problem, involving 3D circulation, day/night cold traps and gravitational settling of haze particles \citep{Parmentier_2013}. Indeed, \citet{Parmentier_2013} predict for HD 209458b that if TiO condenses into particles of size larger than about 1\,\micron\ on the planet's night side, it is plausible that it could be completely depleted from the atmosphere. As TiO and VO begin to condense at temperatures of about 2000\,K \citep[\eg][]{Lodders_2009,Parmentier_2013}, it seems unlikely that TiO/VO will be present at the terminator of HAT-P-32b. Another natural explanation for an underabundance of TiO and VO is a high C/O ratio \citep{Madhusudhan_2012}, or low metallicity, in which case metal hydrides could play a prominent role \citep{Sharp_2007}. We do not construct models including metal hydrides, but assume that their effects would be broadly similar to the metal oxides but consider their presence less plausible.

Several other factors we have yet to consider could contribute to a lack of prominent Na/K features. These includes ionisation which could reduce the abundance of neutral Na/K and therefore the size of the features. This is likely to occur only at high altitudes in the atmosphere, and therefore might reduce the amplitude of the line cores, but cannot easily explain the lack of pressure broadened wings which originate deeper in the atmosphere \citep[\eg][]{Fortney_2003}. Another possibility is that the temperature of the atmosphere at the terminator could be lower than estimated from the equilibrium temperature, therefore reducing the scale height and therefore size of spectral features. However, whilst this could result in diminished Na/K features, this would require an extreme departure from the predicted temperature to explain the lack of observed features alone. Of course, the effects we have discussed are not mutually exclusive, and a combination of them could contribute to a flat transmission spectrum.

We did not attempt to establish which of these atmospheric models provides a better explanation for our observed transmission spectrum in a quantitative manner. To do so robustly would require a full Bayesian analysis; a simple least-squares or maximum likelihood (minimum $\chi^2$) approach would not be satisfactory, as the different families of models being considered have very different levels of complexity. This in turn would require us to chose a sensible parametrisation for each family of models, and suitable priors for each parameter. How to do this is far from obvious, yet the priors would have a major effect on the results, as the flat spectrum means that the data contain relatively little information on the values of many of the parameters. Furthermore, to perform such a model comparison we would have to check for residual correlated errors on the radius ratio measurements obtained from neighbouring wavelength bins, which is also difficult to ascertain. Therefore, we do not perform a statistical model comparison, but rather deduce the relative plausibility in the light of what is currently known about hot Jupiters. Whilst clouds/hazes have been conclusively detected in hot Jupiter atmospheres, TiO or VO have yet to be observed conclusively. Furthermore, the TiO/VO explanation requires an unknown depletion mechanism to remove significant amounts of Ti and V from the atmosphere so that broad TiO/VO features are not detected in the spectrum, yet leave enough to mask Na and K wings, and also with high enough temperatures at the terminator so that gaseous TiO and VO are present. We therefore consider this explanation to be more `fine-tuned', as would a similar explanation involving metal hydrides. This leaves clouds/haze as the more plausible explanation for our spectrum out of those considered, although more precise optical spectra or information at other wavelengths would be required to break this degeneracy.

\begin{table*}
\caption{Transmission spectra of HAT-P-32b for the GP fits to the common mode corrected light curves, including individual fits to the two transits, and a weighted average of the separate fits. The fitted limb darkening parameters are also included.}
\label{tab:trans_spec}
\begin{tabular}{cccccccc}
\hline
\noalign{\smallskip}
\smallskip
Wavelength & \multicolumn{3}{c}{Transit 1} & \multicolumn{3}{c}{Transit 2} & Combined\\ 
\AA & c1 & c2 & ${R_p}/{R_\star}$ & c1 & c2 & ${R_p}/{R_\star}$ & ${R_p}/{R_\star}$\\
\hline
\noalign{\smallskip}
5276 & $0.45\pm0.15$ & $0.33\pm0.20$ & $0.14939\pm0.00397$ & $0.41\pm0.11$ & $0.25\pm0.15$ & $0.15073\pm0.00268$ & $0.15031\pm0.00222$ \\\noalign{\smallskip}
5413 & $0.40\pm0.04$ & $0.36\pm0.07$ & $0.15110\pm0.00077$ & $0.45\pm0.03$ & $0.21\pm0.04$ & $0.15280\pm0.00048$ & $0.15233\pm0.00040$ \\\noalign{\smallskip}
5551 & $0.41\pm0.09$ & $0.26\pm0.13$ & $0.15173\pm0.00186$ & $0.42\pm0.05$ & $0.21\pm0.08$ & $0.15220\pm0.00082$ & $0.15212\pm0.00075$ \\\noalign{\smallskip}
5688 & $0.39\pm0.08$ & $0.24\pm0.12$ & $0.15255\pm0.00177$ & $0.44\pm0.02$ & $0.19\pm0.04$ & $0.15160\pm0.00038$ & $0.15164\pm0.00038$ \\\noalign{\smallskip}
5825 & $0.34\pm0.05$ & $0.31\pm0.07$ & $0.15162\pm0.00125$ & $0.38\pm0.02$ & $0.27\pm0.04$ & $0.15191\pm0.00042$ & $0.15188\pm0.00040$ \\\noalign{\smallskip}
5962 & $0.36\pm0.06$ & $0.31\pm0.10$ & $0.15219\pm0.00108$ & $0.33\pm0.03$ & $0.35\pm0.04$ & $0.15197\pm0.00056$ & $0.15202\pm0.00049$ \\\noalign{\smallskip}
6099 & $0.30\pm0.05$ & $0.38\pm0.08$ & $0.15029\pm0.00099$ & $0.34\pm0.04$ & $0.30\pm0.06$ & $0.15089\pm0.00060$ & $0.15072\pm0.00051$ \\\noalign{\smallskip}
6236 & $0.29\pm0.12$ & $0.41\pm0.18$ & $0.14922\pm0.00228$ & $0.34\pm0.04$ & $0.29\pm0.05$ & $0.15121\pm0.00060$ & $0.15108\pm0.00058$ \\\noalign{\smallskip}
6374 & $0.32\pm0.09$ & $0.32\pm0.14$ & $0.15082\pm0.00153$ & $0.35\pm0.05$ & $0.26\pm0.07$ & $0.15083\pm0.00109$ & $0.15083\pm0.00089$ \\\noalign{\smallskip}
6546 & $0.24\pm0.05$ & $0.35\pm0.08$ & $0.15064\pm0.00081$ & $0.27\pm0.05$ & $0.30\pm0.08$ & $0.15003\pm0.00086$ & $0.15035\pm0.00059$ \\\noalign{\smallskip}
6686 & $0.35\pm0.06$ & $0.18\pm0.09$ & $0.15413\pm0.00108$ & $0.29\pm0.05$ & $0.30\pm0.06$ & $0.15019\pm0.00099$ & $0.15200\pm0.00073$ \\\noalign{\smallskip}
6825 & $0.33\pm0.06$ & $0.33\pm0.09$ & $0.15101\pm0.00098$ & $0.29\pm0.07$ & $0.29\pm0.09$ & $0.15225\pm0.00120$ & $0.15151\pm0.00076$ \\\noalign{\smallskip}
6965 & $0.31\pm0.07$ & $0.28\pm0.10$ & $0.15153\pm0.00109$ & $0.28\pm0.03$ & $0.31\pm0.05$ & $0.15144\pm0.00047$ & $0.15145\pm0.00043$ \\\noalign{\smallskip}
7105 & $0.22\pm0.07$ & $0.43\pm0.10$ & $0.15218\pm0.00126$ & $0.30\pm0.06$ & $0.24\pm0.08$ & $0.15193\pm0.00103$ & $0.15203\pm0.00080$ \\\noalign{\smallskip}
7245 & $0.29\pm0.05$ & $0.31\pm0.08$ & $0.15124\pm0.00070$ & $0.25\pm0.03$ & $0.33\pm0.04$ & $0.15175\pm0.00051$ & $0.15157\pm0.00041$ \\\noalign{\smallskip}
7385 & $0.25\pm0.05$ & $0.35\pm0.07$ & $0.15105\pm0.00071$ & $0.24\pm0.02$ & $0.34\pm0.03$ & $0.15138\pm0.00034$ & $0.15132\pm0.00030$ \\\noalign{\smallskip}
7525 & $0.22\pm0.06$ & $0.32\pm0.08$ & $0.15074\pm0.00095$ & $0.25\pm0.04$ & $0.31\pm0.06$ & $0.15121\pm0.00068$ & $0.15105\pm0.00056$ \\\noalign{\smallskip}
7664 & $0.34\pm0.04$ & $0.27\pm0.06$ & $0.15242\pm0.00089$ & $0.24\pm0.03$ & $0.33\pm0.05$ & $0.15129\pm0.00050$ & $0.15156\pm0.00044$ \\\noalign{\smallskip}
7806 & $0.23\pm0.04$ & $0.34\pm0.06$ & $0.15120\pm0.00073$ & $0.21\pm0.03$ & $0.36\pm0.04$ & $0.15054\pm0.00045$ & $0.15072\pm0.00038$ \\\noalign{\smallskip}
7979 & $0.31\pm0.06$ & $0.20\pm0.09$ & $0.15421\pm0.00108$ & $0.20\pm0.06$ & $0.35\pm0.08$ & $0.15163\pm0.00086$ & $0.15263\pm0.00067$ \\\noalign{\smallskip}
8120 & $0.30\pm0.05$ & $0.19\pm0.07$ & $0.15189\pm0.00075$ & $0.20\pm0.05$ & $0.37\pm0.08$ & $0.15093\pm0.00080$ & $0.15144\pm0.00055$ \\\noalign{\smallskip}
8261 & $0.24\pm0.04$ & $0.26\pm0.07$ & $0.15131\pm0.00074$ & $0.21\pm0.03$ & $0.33\pm0.06$ & $0.15089\pm0.00051$ & $0.15103\pm0.00042$ \\\noalign{\smallskip}
8403 & $0.21\pm0.05$ & $0.31\pm0.07$ & $0.15142\pm0.00103$ & $0.20\pm0.05$ & $0.34\pm0.07$ & $0.15053\pm0.00073$ & $0.15083\pm0.00059$ \\\noalign{\smallskip}
8544 & $0.24\pm0.05$ & $0.32\pm0.07$ & $0.15010\pm0.00081$ & $0.20\pm0.05$ & $0.31\pm0.07$ & $0.15088\pm0.00101$ & $0.15041\pm0.00063$ \\\noalign{\smallskip}
8685 & $0.26\pm0.07$ & $0.22\pm0.11$ & $0.15422\pm0.00114$ & $0.21\pm0.09$ & $0.33\pm0.13$ & $0.15075\pm0.00146$ & $0.15291\pm0.00090$ \\\noalign{\smallskip}
8826 & $0.24\pm0.09$ & $0.26\pm0.14$ & $0.15293\pm0.00156$ & $0.20\pm0.09$ & $0.33\pm0.12$ & $0.15124\pm0.00175$ & $0.15218\pm0.00116$ \\\noalign{\smallskip}
8967 & $0.16\pm0.05$ & $0.36\pm0.08$ & $0.15081\pm0.00091$ & $0.14\pm0.07$ & $0.38\pm0.09$ & $0.15027\pm0.00135$ & $0.15064\pm0.00076$ \\\noalign{\smallskip}
9109 & $0.05\pm0.05$ & $0.44\pm0.08$ & $0.14880\pm0.00182$ & $0.17\pm0.03$ & $0.36\pm0.05$ & $0.15058\pm0.00057$ & $0.15042\pm0.00054$ \\\noalign{\smallskip}
9255 & $0.10\pm0.08$ & $0.22\pm0.10$ & $0.14667\pm0.00415$ & $0.14\pm0.06$ & $0.38\pm0.08$ & $0.15093\pm0.00120$ & $0.15060\pm0.00116$ \\\noalign{\smallskip}
\hline
\end{tabular}
\end{table*}

\section{Discussion}
\label{sect:discussion}

We have presented Gemini-North GMOS observations of the hot Jupiter, HAT-P-32b, during two primary transits. Transmission spectra were extracted for both transits at wavelengths of $520-930$\,\nm\ in $\approx$14\,\nm\ bins, using several comparison stars to perform differential spectro-photometry.

Using the white light curves from the two visits, we derived the system parameters and updated the ephemeris. We found our results to be broadly consistent with the discovery paper \citep{Hartman_2011}, but with larger uncertainties, owing to the variable weather conditions during observations and the increased systematics resulting from extracting light curves from spectra. A previously unknown M-dwarf contaminant star $\approx$2.8$^{\prime\prime}$ from HAT-P-32 was detected in the pre-imaging, which may have diluted the discovery light curves, depending on the apertures used in the data reduction. We therefore caution that some of the system parameters in the discovery paper could be systematically biased; in particular, the planet radius could be marginally larger.

The transmission spectra were extracted using the Gaussian process model of \citet{Gibson_2012} to simultaneously model the transit light curve, instrumental systematics, and time-varying white noise. The flexibility of our GP model proved particularly important for this dataset. Common mode systematics were identified in the spectral light curves, and we used the GP systematics model and residuals from the white light curves to correct for these and significantly improve the precision of our transmission spectra.

The resulting data are consistent with a flat spectrum, and we can rule out broad features larger than about one pressure scale height of the atmosphere. There are two plausible explanations for the absence of the large Na, K, TiO and VO or metal hydride features that we would expect from stellar abundances in HAT-P-32b's atmosphere. The first is the presence of clouds in the upper atmosphere, with large enough particle sizes to act as a grey absorber. The second is that small amounts of TiO/VO (or metal hydrides) could mask the wings of the Na and K lines sufficiently and still be consistent with no large scale variations over about a scale height. The first scenario is more plausible, given that the second requires high enough temperatures at the terminator for gaseous TiO and VO to be present, but with an unknown mechanism required to deplete Ti and V so that large TiO and VO features remain undetected in our spectra, yet leave enough to mask Na and K wings. Furthermore, the cloud scenario naturally explains a flat spectrum, whereas alternative explanations can easily explain the lack of Na/K features, but need to be somewhat `fine-tuned' in order to show no other detectable, broad features in the optical transmission spectrum.
Of course, a combination of both clouds and TiO/VO is also possible, as are contributions from metal hydrides, ionisation, and a smaller scale height if the temperature at the terminator region significantly departs from the equilibrium temperature. In the absence of stronger constraints on the atmosphere of HAT-P-32b, a robust retrieval of detailed information is infeasible. Despite achieving relatively high precision in the transmission spectra, this case highlights the need for broad wavelength coverage to fully understand the complexity of hot Jupiter atmospheres.

\section*{Acknowledgments}

We are grateful to the anonymous referee for careful reading of the manuscript and useful suggestions. J. K. B. acknowledges the support of the John Fell Oxford University Press (OUP) Research Fund for this research. We are extremely grateful for the support provided by the Gemini staff. This work is based on observations obtained at the Gemini Observatory, which is operated by the Association of Universities for Research in Astronomy (AURA) under a cooperative agreement with the NSF on behalf of the Gemini partnership: the National Science Foundation (United States), the Science and Technology Facilities Council (United Kingdom), the National Research Council (Canada), CONICYT (Chile), the Australian Research Council (Australia), CNPq (Brazil) and CONICET (Argentina). 

\bibliography{MyBibliography} 

\begin{thebibliography}{42}
\expandafter\ifx\csname natexlab\endcsname\relax\def\natexlab#1{#1}\fi

\bibitem[{{Barstow} {et~al.}(2013{\natexlab{a}}){Barstow}, {Aigrain}, {Irwin},
  {Bowles}, {Fletcher}, \& {Lee}}]{Barstow_2013a}
{Barstow} J.~K., {Aigrain} S., {Irwin} P.~G.~J., {Bowles} N., {Fletcher} L.~N.,
  {Lee} J.-M., 2013{\natexlab{a}}, \mnras, 430, 1188

\bibitem[{{Barstow} {et~al.}(2013{\natexlab{b}}){Barstow}, {Aigrain}, {Irwin},
  {Fletcher}, \& {Lee}}]{Barstow_2013b}
{Barstow} J.~K., {Aigrain} S., {Irwin} P.~G.~J., {Fletcher} L.~N., {Lee} J.-M.,
  2013{\natexlab{b}}, \mnras, 434, 2616

\bibitem[{{Bean} {et~al.}(2011){Bean}, {D{\'e}sert}, {Kabath}, {Stalder},
  {Seager}, {Miller-Ricci Kempton}, {Berta}, {Homeier}, {Walsh}, \&
  {Seifahrt}}]{Bean_2011}
{Bean} J.~L., {D{\'e}sert} J.-M., {Kabath} P., {Stalder} B., {Seager} S.,
  {Miller-Ricci Kempton} E., {Berta} Z.~K., {Homeier} D., {Walsh} S.,
  {Seifahrt} A., 2011, \apj, 743, 92

\bibitem[{{Bean} {et~al.}(2010){Bean}, {Miller-Ricci Kempton}, \&
  {Homeier}}]{Bean_2010}
{Bean} J.~L., {Miller-Ricci Kempton} E., {Homeier} D., 2010, \nat, 468, 669

\bibitem[{{Berta} {et~al.}(2012){Berta}, {Charbonneau}, {D{\'e}sert},
  {Miller-Ricci Kempton}, {McCullough}, {Burke}, {Fortney}, {Irwin}, {Nutzman},
  \& {Homeier}}]{Berta_2012}
{Berta} Z.~K., {Charbonneau} D., {D{\'e}sert} J.-M., {Miller-Ricci Kempton} E.,
  {McCullough} P.~R., {Burke} C.~J., {Fortney} J.~J., {Irwin} J., {Nutzman} P.,
  {Homeier} D., 2012, \apj, 747, 35

\bibitem[{{Bishop}(2006)}]{Bishop}
{Bishop} C.~M., 2006, {Pattern Recognition and Machine Learning}. {Springer}

\bibitem[{{Brown}(2001)}]{Brown_2001}
{Brown} T.~M., 2001, \apj, 553, 1006

\bibitem[{{Burrows} {et~al.}(2000){Burrows}, {Marley}, \&
  {Sharp}}]{Burrows_2000}
{Burrows} A., {Marley} M.~S., {Sharp} C.~M., 2000, \apj, 531, 438

\bibitem[{{Carter} \& {Winn}(2009)}]{Carter_2009}
{Carter} J.~A., {Winn} J.~N., 2009, \apj, 704, 51

\bibitem[{{Charbonneau} {et~al.}(2002){Charbonneau}, {Brown}, {Noyes}, \&
  {Gilliland}}]{Charbonneau_2002}
{Charbonneau} D., {Brown} T.~M., {Noyes} R.~W., {Gilliland} R.~L., 2002, \apj,
  568, 377

\bibitem[{{Claret}(2004)}]{Claret_2004}
{Claret} A., 2004, \aap, 428, 1001

\bibitem[{{Crossfield} {et~al.}(2013){Crossfield}, {Barman}, {Hansen}, \&
  {Howard}}]{Crossfield_2013}
{Crossfield} I.~J.~M., {Barman} T., {Hansen} B.~M.~S., {Howard} A.~W., 2013,
  arxiv:1308.6580

\bibitem[{{Crouzet} {et~al.}(2012){Crouzet}, {McCullough}, {Burke}, \&
  {Long}}]{Crouzet_2012}
{Crouzet} N., {McCullough} P.~R., {Burke} C., {Long} D., 2012, \apj, 761, 7

\bibitem[{{Deming} {et~al.}(2013){Deming}, {Wilkins}, {McCullough}, {Burrows},
  {Fortney}, {Agol}, {Dobbs-Dixon}, {Madhusudhan}, {Crouzet}, {Desert},
  {Gilliland}, {Haynes}, {Knutson}, {Line}, {Magic}, {Mandell}, {Ranjan},
  {Charbonneau}, {Clampin}, {Seager}, \& {Showman}}]{Deming_2013}
{Deming} D., {Wilkins} A., {McCullough} P., {Burrows} A., {Fortney} J.~J.,
  {Agol} E., {Dobbs-Dixon} I., {Madhusudhan} N., {Crouzet} N., {Desert} J.-M.,
  {Gilliland} R.~L., {Haynes} K., {Knutson} H.~A., {Line} M., {Magic} Z.,
  {Mandell} A.~M., {Ranjan} S., {Charbonneau} D., {Clampin} M., {Seager} S.,
  {Showman} A.~P., 2013, \apj, 774, 95

\bibitem[{{D{\'e}sert} {et~al.}(2008){D{\'e}sert}, {Vidal-Madjar}, {Lecavelier
  Des Etangs}, {Sing}, {Ehrenreich}, {H{\'e}brard}, \& {Ferlet}}]{Desert_2008}
{D{\'e}sert} J.-M., {Vidal-Madjar} A., {Lecavelier Des Etangs} A., {Sing} D.,
  {Ehrenreich} D., {H{\'e}brard} G., {Ferlet} R., 2008, \aap, 492, 585

\bibitem[{{Fortney} {et~al.}(2008){Fortney}, {Lodders}, {Marley}, \&
  {Freedman}}]{Fortney_2008}
{Fortney} J.~J., {Lodders} K., {Marley} M.~S., {Freedman} R.~S., 2008, \apj,
  678, 1419

\bibitem[{{Fortney} {et~al.}(2003){Fortney}, {Sudarsky}, {Hubeny}, {Cooper},
  {Hubbard}, {Burrows}, \& {Lunine}}]{Fortney_2003}
{Fortney} J.~J., {Sudarsky} D., {Hubeny} I., {Cooper} C.~S., {Hubbard} W.~B.,
  {Burrows} A., {Lunine} J.~I., 2003, \apj, 589, 615

\bibitem[{{Gibson} {et~al.}(2013){Gibson}, {Aigrain}, {Barstow}, {Evans},
  {Fletcher}, \& {Irwin}}]{Gibson_2013}
{Gibson} N.~P., {Aigrain} S., {Barstow} J.~K., {Evans} T.~M., {Fletcher} L.~N.,
  {Irwin} P.~G.~J., 2013, \mnras, 428, 3680

\bibitem[{{Gibson} {et~al.}(2010){Gibson}, {Aigrain}, {Pollacco}, {Barros},
  {Hebb}, {Hrudkov{\'a}}, {Simpson}, {Skillen}, \& {West}}]{Gibson_2010b}
{Gibson} N.~P., {Aigrain} S., {Pollacco} D.~L., {Barros} S.~C.~C., {Hebb} L.,
  {Hrudkov{\'a}} M., {Simpson} E.~K., {Skillen} I., {West} R., 2010, \mnras,
  404, L114

\bibitem[{{Gibson} {et~al.}(2012{\natexlab{a}}){Gibson}, {Aigrain}, {Pont},
  {Sing}, {D{\'e}sert}, {Evans}, {Henry}, {Husnoo}, \&
  {Knutson}}]{Gibson_2012b}
{Gibson} N.~P., {Aigrain} S., {Pont} F., {Sing} D.~K., {D{\'e}sert} J.-M.,
  {Evans} T.~M., {Henry} G., {Husnoo} N., {Knutson} H., 2012{\natexlab{a}},
  \mnras, 422, 753

\bibitem[{{Gibson} {et~al.}(2012{\natexlab{b}}){Gibson}, {Aigrain}, {Roberts},
  {Evans}, {Osborne}, \& {Pont}}]{Gibson_2012}
{Gibson} N.~P., {Aigrain} S., {Roberts} S., {Evans} T.~M., {Osborne} M., {Pont}
  F., 2012{\natexlab{b}}, \mnras, 419, 2683

\bibitem[{{Gibson} {et~al.}(2011){Gibson}, {Pont}, \& {Aigrain}}]{Gibson_2011}
{Gibson} N.~P., {Pont} F., {Aigrain} S., 2011, \mnras, 411, 2199

\bibitem[{{Hartman} {et~al.}(2011){Hartman}, {Bakos}, {Torres}, {Latham},
  {Kov{\'a}cs}, {B{\'e}ky}, {Quinn}, {Mazeh}, {Shporer}, {Marcy}, {Howard},
  {Fischer}, {Johnson}, {Esquerdo}, {Noyes}, {Sasselov}, {Stefanik},
  {Fernandez}, {Szklen{\'a}r}, {L{\'a}z{\'a}r}, {Papp}, \&
  {S{\'a}ri}}]{Hartman_2011}
{Hartman} J.~D., {Bakos} G.~{\'A}., {Torres} G., {Latham} D.~W., {Kov{\'a}cs}
  G., {B{\'e}ky} B., {Quinn} S.~N., {Mazeh} T., {Shporer} A., {Marcy} G.~W.,
  {Howard} A.~W., {Fischer} D.~A., {Johnson} J.~A., {Esquerdo} G.~A., {Noyes}
  R.~W., {Sasselov} D.~D., {Stefanik} R.~P., {Fernandez} J.~M., {Szklen{\'a}r}
  T., {L{\'a}z{\'a}r} J., {Papp} I., {S{\'a}ri} P., 2011, \apj, 742, 59

\bibitem[{{Hook} {et~al.}(2004){Hook}, {J{\o}rgensen}, {Allington-Smith},
  {Davies}, {Metcalfe}, {Murowinski}, \& {Crampton}}]{Hook_2004}
{Hook} I.~M., {J{\o}rgensen} I., {Allington-Smith} J.~R., {Davies} R.~L.,
  {Metcalfe} N., {Murowinski} R.~G., {Crampton} D., 2004, \pasp, 116, 425

\bibitem[{{Huitson} {et~al.}(2012){Huitson}, {Sing}, {Vidal-Madjar},
  {Ballester}, {Lecavelier des Etangs}, {D{\'e}sert}, \& {Pont}}]{Huitson_2012}
{Huitson} C.~M., {Sing} D.~K., {Vidal-Madjar} A., {Ballester} G.~E.,
  {Lecavelier des Etangs} A., {D{\'e}sert} J.-M., {Pont} F., 2012, \mnras, 422,
  2477

\bibitem[{{Irwin} {et~al.}(2008){Irwin}, {Teanby}, {de Kok}, {Fletcher},
  {Howett}, {Tsang}, {Wilson}, {Calcutt}, {Nixon}, \& {Parrish}}]{Irwin_2008}
{Irwin} P.~G.~J., {Teanby} N.~A., {de Kok} R., {Fletcher} L.~N., {Howett}
  C.~J.~A., {Tsang} C.~C.~C., {Wilson} C.~F., {Calcutt} S.~B., {Nixon} C.~A.,
  {Parrish} P.~D., 2008, J. Quant. Spectrosc. Radiat. Transfer, 109, 1136

\bibitem[{{Lee} {et~al.}(2012){Lee}, {Fletcher}, \& {Irwin}}]{Lee_2012}
{Lee} J.-M., {Fletcher} L.~N., {Irwin} P.~G.~J., 2012, \mnras, 420, 170

\bibitem[{Lodders(2010)}]{Lodders_2009}
Lodders K., 2010, Exoplanet Chemistry, Wiley-VCH Verlag GmbH \& Co. KGaA, pp.
  157--186

\bibitem[{{Madhusudhan}(2012)}]{Madhusudhan_2012}
{Madhusudhan} N., 2012, \apj, 758, 36

\bibitem[{{Mandel} \& {Agol}(2002)}]{mandel_agol_2002}
{Mandel} K., {Agol} E., 2002, \apjl, 580, L171

\bibitem[{{Parmentier} {et~al.}(2013){Parmentier}, {Showman}, \&
  {Lian}}]{Parmentier_2013}
{Parmentier} V., {Showman} A.~P., {Lian} Y., 2013, arxiv:1301.4522

\bibitem[{{Pont} {et~al.}(2008){Pont}, {Knutson}, {Gilliland}, {Moutou}, \&
  {Charbonneau}}]{Pont_2008}
{Pont} F., {Knutson} H., {Gilliland} R.~L., {Moutou} C., {Charbonneau} D.,
  2008, \mnras, 385, 109

\bibitem[{{Pont} {et~al.}(2013){Pont}, {Sing}, {Gibson}, {Aigrain}, {Henry}, \&
  {Husnoo}}]{Pont_2013}
{Pont} F., {Sing} D.~K., {Gibson} N.~P., {Aigrain} S., {Henry} G., {Husnoo} N.,
  2013, \mnras, 432, 2917

\bibitem[{{Pont} {et~al.}(2006){Pont}, {Zucker}, \& {Queloz}}]{Pont_2006}
{Pont} F., {Zucker} S., {Queloz} D., 2006, \mnras, 373, 231

\bibitem[{Schwarz(1978)}]{Schwarz_1978}
Schwarz G., 1978, Ann. Statist., 6, 461

\bibitem[{{Seager} \& {Sasselov}(2000)}]{Seager_2000}
{Seager} S., {Sasselov} D.~D., 2000, \apj, 537, 916

\bibitem[{{Sharp} \& {Burrows}(2007)}]{Sharp_2007}
{Sharp} C.~M., {Burrows} A., 2007, \apjs, 168, 140

\bibitem[{{Sing} {et~al.}(2012){Sing}, {Huitson}, {Lopez-Morales}, {Pont},
  {D{\'e}sert}, {Ehrenreich}, {Wilson}, {Ballester}, {Fortney}, {Lecavelier des
  Etangs}, \& {Vidal-Madjar}}]{Sing_2012}
{Sing} D.~K., {Huitson} C.~M., {Lopez-Morales} M., {Pont} F., {D{\'e}sert}
  J.-M., {Ehrenreich} D., {Wilson} P.~A., {Ballester} G.~E., {Fortney} J.~J.,
  {Lecavelier des Etangs} A., {Vidal-Madjar} A., 2012, \mnras, 426, 1663

\bibitem[{{Sing} {et~al.}(2011){Sing}, {Pont}, {Aigrain}, {Charbonneau},
  {D{\'e}sert}, {Gibson}, {Gilliland}, {Hayek}, {Henry}, {Knutson}, {Lecavelier
  Des Etangs}, {Mazeh}, \& {Shporer}}]{Sing_2011}
{Sing} D.~K., {Pont} F., {Aigrain} S., {Charbonneau} D., {D{\'e}sert} J.-M.,
  {Gibson} N., {Gilliland} R., {Hayek} W., {Henry} G., {Knutson} H.,
  {Lecavelier Des Etangs} A., {Mazeh} T., {Shporer} A., 2011, \mnras, 416, 1443

\bibitem[{{Sing} {et~al.}(2008){Sing}, {Vidal-Madjar}, {D{\'e}sert},
  {Lecavelier des Etangs}, \& {Ballester}}]{Sing_2008}
{Sing} D.~K., {Vidal-Madjar} A., {D{\'e}sert} J.-M., {Lecavelier des Etangs}
  A., {Ballester} G., 2008, \apj, 686, 658

\bibitem[{{Stevenson} {et~al.}(2013){Stevenson}, {Bean}, {Seifahrt}, {Desert},
  {Madhusudhan}, {Bergmann}, {Kreidberg}, \& {Homeier}}]{Stevenson_2013}
{Stevenson} K.~B., {Bean} J.~L., {Seifahrt} A., {Desert} J.-M., {Madhusudhan}
  N., {Bergmann} M., {Kreidberg} L., {Homeier} D., 2013, arxiv:1305.1670

\bibitem[{{Waldmann}(2012)}]{Waldmann_2012}
{Waldmann} I.~P., 2012, \apj, 747, 12

\end{thebibliography}
\bibliographystyle{mn2e_astronat} 

\label{lastpage}

\end{document}